\documentclass[10pt]{article}
\pdfoutput=1

\usepackage{amsmath}
\usepackage{amssymb}

\usepackage{graphicx}

\usepackage{cite}

\usepackage{color} 

\usepackage{setspace} 

\usepackage[labelfont=bf,labelsep=period,justification=raggedright]{caption}

\makeatletter
\renewcommand{\@biblabel}[1]{\quad#1.}
\makeatother

\date{}

\usepackage[usenames,dvipsnames]{xcolor}
\usepackage{pdfpages}

\usepackage[T1]{fontenc}

\usepackage{empheq}
\newcommand*\myyelbox[1]{%
\fcolorbox{black}{yellow}{\hspace{1em}#1\hspace{1em}}}

\usepackage{titlesec}
\titleformat*{\section}{\normalfont\Large\bfseries\color{Blue}}
\titleformat*{\subsection}{\normalfont\large\bfseries\color{MidnightBlue}}
\titleformat*{\subsubsection}{\normalfont\normalsize\bfseries\color{Turquoise}}

\usepackage{enumerate}
\usepackage[colorlinks=true, linkcolor=NavyBlue, citecolor=Red, urlcolor=NavyBlue, linktoc=page]{hyperref}

\begin{document}

\begin{flushleft}
{\Large
\textbf{A mathematical model for distinguishing bias from sensitivity effects in multialternative detection tasks}
}
\linebreak
\\
Devarajan Sridharan$^{1,\ast}$, 
Nicholas A Steinmetz$^{1,2}$, 
Tirin Moore$^{1,3}$,
Eric I Knudsen$^{1}$ 
\linebreak
\\
\bf{1} Department of Neurobiology, Stanford University School of Medicine, Stanford, California, USA
\\
\bf{2} Program in Neurosciences, Stanford University School of Medicine, Stanford, California, USA
\\
\bf{3} Howard Hughes Medical Institute, Stanford University School of Medicine, Stanford, California, USA 
\\
$\ast$ E-mail: dsridhar@stanford.edu
\end{flushleft}

\section*{Abstract}
Studies investigating the neural bases of cognitive phenomena such as perception, attention and decision-making increasingly employ multialternative task designs. It is essential in such designs to distinguish the neural correlates of behavioral choices arising from changes in perceptual factors, such as enhanced sensitivity to sensory information, from those arising from changes in decisional factors, such as a stronger bias for a particular response or choice (choice bias). To date, such a distinction is not possible with established approaches. Thus, there is a critical need for a theoretical approach that distinguishes the effects of changes in sensitivity from those of changes in choice bias in multialternative tasks. 

Here, we introduce a mathematical model that decouples choice bias from perceptual sensitivity effects in multialternative detection tasks: multialternative tasks that incorporate catch trials to measure the ability to detect one among multiple (potential) stimuli or stimulus features. By formulating the perceptual decision in a novel, multidimensional signal detection framework, our model identifies the distinct effects of bias and sensitivity on behavioral choices. With a combination of analytical and numerical approaches, we demonstrate that model parameters (sensitivity, bias) are estimated reliably and uniquely, even in tasks involving arbitrarily large numbers of alternatives. 

Model simulations revealed that ignoring choice bias or performance during catch trials produced systematically inaccurate estimates of perceptual sensitivity, a finding that has important implications for interpreting behavioral data in multialternative detection and cued attention tasks. The model will find important application in identifying the effects of neural perturbations (stimulation or inactivation) on an animal's perception in multialternative attention and decision-making tasks.

\clearpage
\section*{Author Summary}
There is increasing interest in measuring the neural basis of perception and cognition with multialternative tasks (tasks with more than two stimulus or response alternatives). Choice biases represent tendencies (idiosyncratic or rational) to favor responses to one or a few alternatives over others. Thus, an important challenge when interpreting behavioral data in such multialternative tasks is to infer which aspects of behavior were driven by changes in the subject's perception, and which by choice biases.

We have developed a mathematical model that overcomes this challenge. The model decouples the effects of choice bias when estimating perceptual sensitivity in widely-used, multialternative detection tasks: tasks that measure the ability to detect one among several potential stimuli or stimulus features. We provide numerical methods for estimating choice bias from multialternative behavioral data, and highlight key caveats for interpreting behavioral data without accounting for such bias effects. Our model provides a powerful and widely applicable tool for analyzing the contribution of bias to behavioral choices in human and animal studies of perception, attention and decision-making.

\clearpage
{\color{black}
\tableofcontents
}


\clearpage
\section*{Introduction}
\addcontentsline{toc}{section}{Introduction}

Decisions in the real world involve making a categorical judgment or choice based on careful evaluation of noisy sensory evidence. Besides sensory evidence, behavioral biases contribute importantly to the decision-making process \cite{goldshadlen, goldbennur, mcsdtbk}. Biases may reflect an innate preference for a specific choice that manifests, for instance, as an idiosyncratic tendency for selecting one choice among many equally likely alternatives \cite{klein2001, goldbennur}. Conversely, biases may be rapidly and reversibly induced with specific task manipulations. For instance, cueing the location of an upcoming stimulus, either explicitly with a spatial cue or implicitly by temporarily increasing the frequency of presentation at a particular location, could result in the subject developing a bias for selecting that location over other locations in the time-span of a few trials \cite{carpenter, hanksshadlen, mulder}. Systematic biases for specific choices (``choice biases'') confound the ability to evaluate the subject's sensitivity to sensory evidence. Hence, in studies of human and animal behavior much effort is invested in the careful development of experimental designs and training protocols that minimize, or train away, biases, although this approach may not always be practical. 

Theoretical frameworks provide a complementary approach for accounting for choice bias: they quantify it. Such frameworks are based on a testable model of the decision-making process, and permit principled, quantitative estimation of the contribution of choice bias to the subject's responses. Among such theoretical frameworks, signal detection theory (SDT) is a simple, but powerful, decision-making framework for accounting for choice bias in binary choice tasks, such as the two-alternative forced choice (2-AFC) or Yes/No detection tasks \cite{greenswetsbk, mcsdtbk}. 

In two-alternative (Yes/No) detection tasks, the experimenter seeks to measure a (human or animal) subject's perceptual sensitivity to detect a stimulus or a stimulus feature in the display. The subject is presented with a series of behavioral trials; the stimulus (or stimulus feature) is presented on a random subset of these trials and is absent in others. When the stimulus is detected the subject reports it with a ``Yes'' response; otherwise, the subject reports a ``No'' response. Signal detection theory models the subject's perceptual decision as an inherently noisy process. In the SDT framework for the binary choice (Yes/No) task, the subject decides between the two, mutually exclusive events (was the stimulus present, or not?) by weighing the sensory evidence for each (Figure 2A). The decision is based on a latent random variable, the decision variable ($\Psi$), whose mean depends on the strength of the sensory evidence, and whose variance arises from the noisiness of the sensory evidence across trials \cite{greenswetsbk}. On trials in which the decision variable ($\Psi$) falls above a cutoff value ($c$, Figure 2A, red vertical line), the subject reports having detected the stimulus (``Yes''). 

The cutoff value or ``choice criterion'' ($c$) represents the subject's bias for choosing to report one event over the other. When the subject is highly biased toward the ``Yes'' choice, she/he(/it) adopts a low value for the choice criterion which manifests as a tendency to guess at detection, i.e., to report having detected the stimulus even when no stimulus was presented (a high rate of ``false-alarms''). Conversely, when the subject is highly biased toward the ``No'' choice, she/he adopts a high criterion which manifests as a conservative tendency to not report detection even on trials when the stimulus was presented (a high rate of ``misses''). Having accounted for the subject's tendency to guess (bias), the subject's ``perceptual sensitivity'' (denoted $d$) to detect the stimulus is analytically inferred from the proportion of false-alarms and misses based on assumptions about the nature of the decision variable distribution \cite{greenswetsbk}. 

Now, consider the following scenario: an experimenter seeks to measure a subject's perceptual sensitivity for detecting stimuli at not one, but two different locations (Figure 1A). To accomplish this, let the experimenter employ a two-alternative forced choice design (2-AFC), so that instead of the ``No'' response, the subject reports one of the two locations at which the stimulus was presented (Figure 1A, upper). 2-AFC designs (and m-alternative forced-choice designs, in general) present an important limitation for measuring detection sensitivity in such tasks: the potential confound introduced by a ``guessing'' strategy. For instance, if the subject were highly sensitive to detecting the stimulus at one of the two locations (say location 1), she/he(/it) could use the strategy of ``guessing'' that the stimulus was presented at location 2 when no stimulus was detected at location 1. In this sense, the subject's responses are not indicative of detecting the stimulus at location 2 and, hence, cannot be used to estimate perceptual sensitivity for detection at location 2. 

To overcome this limitation, in addition to offering two or more stimulus alternatives, experimental designs include ``catch'' (stimulus absent) trials \cite{cavanaughwurtz, cohenmaunsell, moorefallah, raymaunsell, zenonkrauzlis}. In such trials, no stimulus is presented or no actionable sensory evidence is available, and the subject is rewarded for not making a response (commonly called NoGo, analogous to the ``No'' response in the Yes/No task; Figure 1A, lower). Measuring false-alarm rates (the proportion of guesses) during such catch trials identifies and controls for a subject's guessing strategy. Thus, catch trials are, in general, necessary for multialternative tasks that seek to measure detection sensitivity at multiple spatial locations or to multiple potential values of a stimulus attribute (e.g., to different colors or stimulus orientations). We term such a task for measuring detection performance with multiple stimulus/response alternatives and catch trials a ``multiple alternative (or multialternative) detection task''. 

Despite its considerable success with accounting for choice bias in 2-AFC and Yes/No tasks, conventional SDT cannot readily be applied to tasks with more than two alternatives. Yet, accounting for bias in such multialternative detection tasks (with catch trials) remains an important, open problem \cite{mcsdt10} (pp.250-251) \cite{decarlo}. Here, we introduce a multidimensional signal detection model to decouple choice bias from perceptual sensitivity in multialternative detection tasks. We develop analytical and numerical approaches for estimating perceptual (detection) sensitivity and choice criteria from measured response probabilities, and highlight key caveats with ignoring choice bias when interpreting behavior in multialternative detection tasks. Our model provides a general and powerful framework for inferring an animal's perception from its behavior in multialternative attention and decision-making tasks.

\section*{Results}
\addcontentsline{toc}{section}{Results}

\subsection*{The multialternative detection task: motivation for a multidimensional model}
\addcontentsline{toc}{subsection}{The multialternative detection task: motivation for a multidimensional model}

\begin{figure}[!t]
\begin{center}
\includegraphics[width=6.25in]{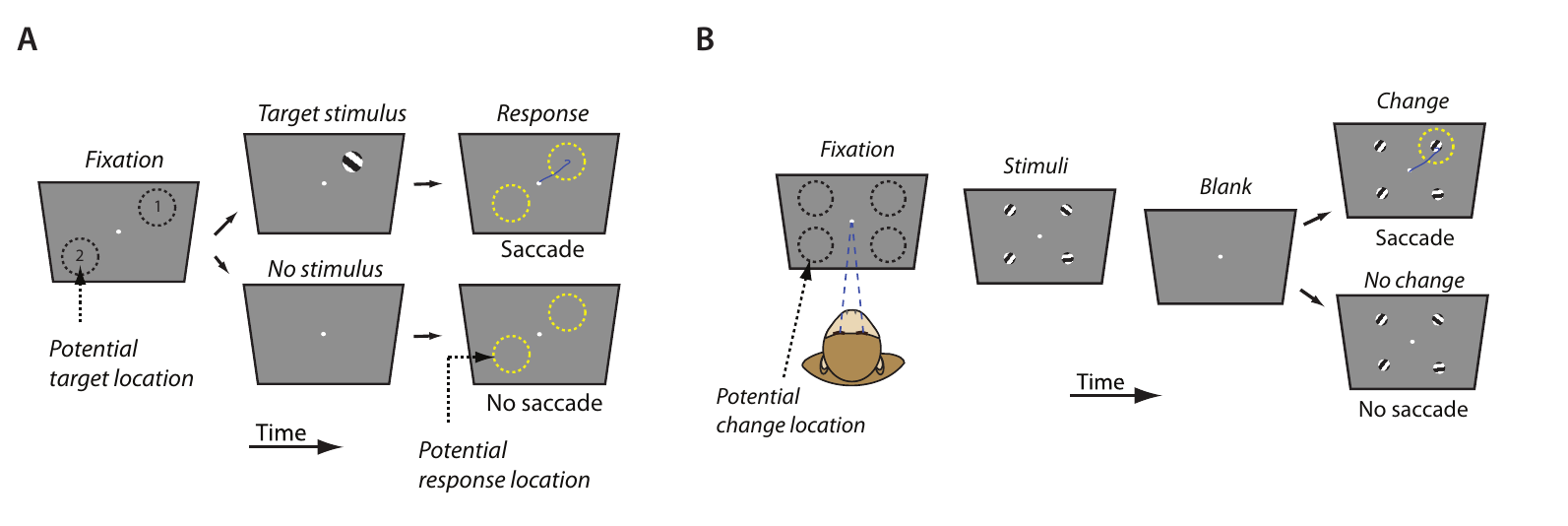}
\end{center}
\caption{
{\bf Multialternative detection task.} 
({\bf A}) Two-alternative detection (2-ADC) task. The subject initiates a trial by fixating on a zeroing dot. On some trials (``stimulus'' trials, upper sequence), a target stimulus (here, a grating) is briefly presented at one of two potential locations (dashed black circles) on the screen. The subject is rewarded for detecting and indicating the location of the target with a saccade (blue line, ``Go'' response) to the appropriate response box (dashed yellow circles). On other trials, (``catch'' trials, lower sequence), no target is presented for a prolonged period following fixation. In these trials, the subject is rewarded for maintaining fixation on the zeroing dot (``NoGo'' response) following the appearance of the response boxes. 
({\bf B}) Multialternative \textit{change} detection (m-ADC) task. Following fixation of a central dot, the subject is presented with m (here, m=4) oriented gratings. At a random time following stimulus onset, the display goes blank briefly ($\sim$300 ms). Then, the four stimuli reappear. On most of the trials, one of the four gratings has changed in orientation (change trials), and on the remaining trials none of the stimuli changed (catch trials). The subject is rewarded for saccading to the location of the change (change trials) or for maintaining fixation on trials when no change occurred (catch trials).
}
\label{fig_task}
\end{figure}

We develop a multidimensional signal detection model to decouple choice bias from perceptual sensitivity in multialternative detection tasks (with catch trials). For brevity, we will refer to such tasks as \textit{m-ADC} tasks. To facilitate development of the model we choose a particular kind of simple detection task, a multiple alternative spatial detection task (Figure 1A). The theory is applicable to other kinds of task designs as well, such as those involving the detection of a \textit{change} in a stimulus feature (e.g. orientation; Figure 1B) employed by previous studies \cite{cavanaughwurtz, cohenmaunsell, raymaunsell}.

In this spatial detection m-ADC task, a subject is presented with a briefly-flashed stimulus that can occur at one (or none) of several potential spatial locations (Figure 1A). Trials in which a stimulus was presented are termed ``stimulus trials'', whereas trials in which no stimulus was presented are termed ``catch trials''. The subject reports the location at which s/he perceived the stimulus, for instance with a saccadic eye movement (Figure 1A, top sequence). In case no stimulus was detected (e.g. during catch trials), the subject is rewarded for making a NoGo response (Figure 1A, bottom sequence). Incorporating catch trials and NoGo responses makes the m-ADC task a more general version of multialternative forced-choice (m-AFC) task.

To motivate the development of the multidimensional model for the m-ADC task, we first demonstrate why the task cannot be modeled with multiple one-dimensional binary choice models. First, consider a binary choice (Yes/No) spatial detection task (Figure 2A). In this task, the stimulus is presented either at a location (location 1) or not at all (catch). Conventional SDT models the binary (Yes/No) decision as a process of selecting one of two hypotheses ($h_0$: no stimulus or $h_1$: stimulus at location 1) based on their relative likelihoods, given the sensory evidence. The decision variable ($\Psi$), proportional to the likelihood ratio of the hypotheses, is modeled as a Gaussian random variable with unit variance. The subject chooses $h_1$ if the decision variable exceeds a particular value, the \textsl{criterion}; such a specification permits maximizing accuracy by scaling the value of the criterion to match the ratio of the priors (for a detailed discussion, see \cite{greenswetsbk}). The well-known 2 $\times$ 2 contingency table for this type of task is shown in Table S1A. 

Based on the proportion of hits ($\mathrm{HR}$) -- the proportion of trials in which the subject correctly reported a detection when a stimulus was presented -- and false-alarms ($\mathrm{FA}$) -- the proportion of trials in which the subject incorrectly reported a detection when no stimulus was presented -- SDT provides a simple, one-dimensional formalism for estimating the subject's ``perceptual sensitivity'' and ``choice criterion''. Perceptual sensitivity, $d=\Phi^{-1}({HR}) - \Phi^{-1}({FA})$, represents a measure of the discriminability (or overlap) of the decision variable distributions for the two hypotheses, and the ``choice criterion'', $c=\Phi^{-1}({FA})$, a measure of bias for choosing one hypothesis over the other (where $\Phi^{-1}$ represents the \textsl{probit} function, the inverse cumulative distribution function associated with the standard normal distribution). 

Let us now consider the 2-ADC task where the stimulus can be presented at one of two locations, in addition to not being presented at all (Figure 1A). The 3x3 contingency table for this task is shown in Table S1B. For this task, the decision must be made among three hypotheses: $h_1$, stimulus at location 1; $h_2$, stimulus at location 2; or $h_0$, no stimulus at either location. 

Such a two-alternative detection task cannot be modeled as two Yes/No detection tasks executed in parallel with two independent binary choice detection models. To illustrate why, let us attempt to model the responses in such a task with two one-dimensional SDT models, one for each of the two locations (Figure 2B). As shown in the figure, the ``noise'' $(\Psi|h_0)$ distribution at each location (black) is modeled as a unit-variance Gaussian that represents the distribution of the decision variable during no-stimulus (catch) trials. The ``signal'' distribution at each location k, $(\Psi|h_k)$ corresponds to a translated version of the noise distribution (with a higher mean value) at that location. The perceptual sensitivity at each location k, $d_k$, is quantified as the difference between the means of the signal and noise distributions at that location (for noise standard deviation of unity; see next section for details). On a given trial, the subject reports a Go response at location k if the decision variable at that location exceeds the choice criterion, $c_k$; if the decision variable does not exceed the choice criterion at either location, the subject reports a NoGo response.

We now estimate perceptual sensitivity and choice criterion for each location with this pair of independent, one-dimensional models. As per this model, the false alarm rate at, say, location 1 ($\mathrm{FA}_1$, the probability of incorrectly reporting a stimulus at location 1 when no stimulus was presented; hatched area in Figure 2B) determines the choice criterion ($c_1$) at this location ($c_1=\Phi^{-1}({FA}_1)$, as before). Similarly, the hit rate at location 1 ($\mathrm{HR}_1$, the probability of correctly reporting a stimulus at location 1 when it was presented at that location, red shaded area in Figure 2B), in conjunction with the false-alarm rate (${FA}_1$), determines the perceptual sensitivity ($d_1$) at this location ($d_1=\Phi^{-1}({HR}_1) - \Phi^{-1}({FA}_1)$). The criterion and sensitivity at location 2 may be similarly estimated based on the hit-rate (${HR}_2$) and false-alarm rate (${FA}_2$) at location 2. Note that when analyzed in this way (as a pair of one-dimensional models), parameters (d and c) at each location are estimated independently of responses at the other location. 

Although intuitively appealing, such a model has the following major drawbacks. First, under this model, the miss-rate at location k (the probability of giving a NoGo response when a stimulus was presented at location k) is exactly the complement of the hit rate at location k (${MR}_k = 1 - \mathrm{HR}_k$). Thus, the model cannot deal with the scenario in which the animal responds to a location that is different from the location at which the stimulus was presented (shaded squares, Table S1B). Second, such a model duplicates the specification of the correct-rejection rate, the complement of the false-alarm rate at each location (${CR}_k = 1-\mathrm{FA}_k$, grey shaded area, Figure 2B), whereas the observed correct rejection rate (probability of giving a NoGo response when no stimulus was presented at either location) is, by definition, a unique quantity. Third, the model specifies that the animal will respond to the location at which the decision variable exceeds criterion. What if decision variables were to exceed criteria at both locations on a particular trial? This causes a decision conflict in the model, because responses cannot be made to more than one location on a given trial. These drawbacks demonstrate that, two-alternative (or, in general, multialternative) detection tasks cannot be analyzed with two (or multiple) independent one-dimensional binary choice signal detection models.

\subsection*{Two-dimensional signal detection model for the two-alternative detection (2-ADC) task}
\addcontentsline{toc}{subsection}{Two-dimensional signal detection model for the two-alternative detection (2-ADC) task}

We develop a multidimensional model, first, for a two alternative detection task (Figure 1A) and in a later section generalize the model to a task with several alternatives ($m>2$). We briefly describe the model below, and provide a detailed analytical formulation in later sections.

In our two-dimensional signal detection model, independent decision variables represent stimuli at each location along orthogonal perceptual dimensions (Figure 2C). When no stimulus is presented (catch trials), the joint distribution of decision variables (the ``noise'' distribution) is centered at the origin, with equal variance along each perceptual dimension (Figure 2C, black). A stimulus presented at a particular location translates the noise distribution along the perceptual dimension (decision variable axis) for that location (the ``signal'' distribution; Figure 2C, red or blue). The magnitude of this translation is inversely related to the overlap between the signal and noise distributions, and is defined as the perceptual sensitivity ($d_k$) for each location, $k$.

The model posits that, while choosing a response, the subject employs an independent choice criterion ($c_k$) for each location: on each trial a response is made to the location at which the decision variable exceeds the choice criterion. A difference in criteria between the two locations gives rise to a choice bias (relative preference) for one location over the other. If decision variables at both locations exceed their respective criteria, the response is made to the location at which the difference between the decision variable and the corresponding choice criterion was the greatest. If no decision variable exceeds its respective criterion, the subject abstains from responding (or gives a NoGo response; Figure 2C, Y=0, gray shaded region). Under this model, the response probability for each stimulus contingency is directly related to the proportion (integral) of the corresponding joint distribution within its decision boundary.

This multidimensional 2-ADC model overcomes the drawbacks of the (aforementioned) multiple binary choice model formulation (Figure 2B). First, this model captures the possibility that the subject could respond to a location that is different from the stimulus location (Table S1B, shaded cells); the probability of such a response is given by the proportion of the signal distribution for a stimulus at a location (e.g., Figure 2C; red distribution) that lies within the decision boundary for a response at a different location (e.g., Y=2 decision boundary). Second, the probability of a correct rejection is uniquely specified; this corresponds to the proportion of the noise distribution that lies within the decision boundary of the NoGo response (Figure 2C; gray shaded area). Third, the decision rule specifies a clear response for the scenario in which the decision variable exceeds criterion at both locations. Our formulation of the model that incorporates catch trials and NoGo responses is more general and subsumes previous models for multialternative forced-choice tasks that do not (Appendix B, Supporting Information).

\begin{figure}[!t]
\begin{center}
\includegraphics[width=6.25in]{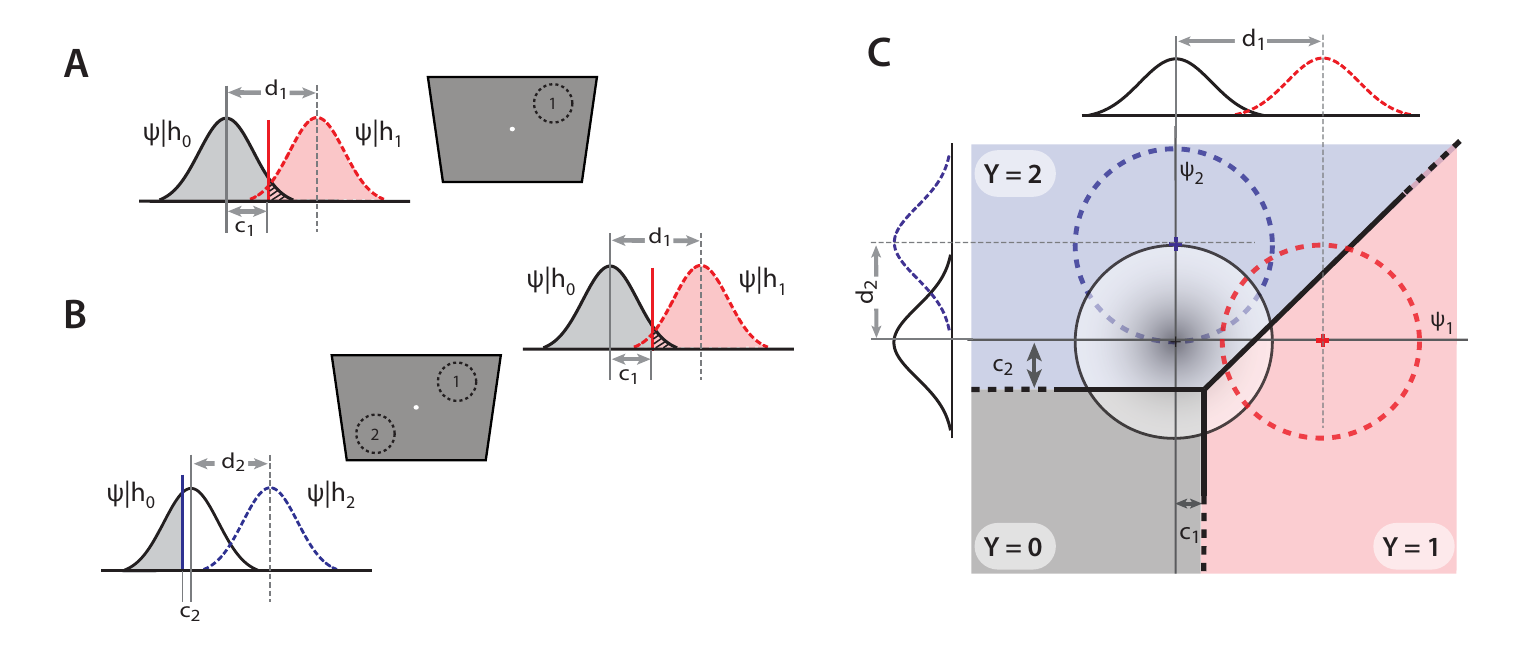}
\end{center}
\caption{
{\bf Signal detection models for the multialternative detection task.} 
({\bf A}) A two-alternative forced choice (Yes/No) task modeled with a one-dimensional signal detection model. Black Gaussian: noise distribution: red Gaussian: signal distribution; red shading: hit-rate; gray shading: correct rejection rate; hatched region: false-alarm rate; $d_1$: perceptual sensitivity for detecting the stimulus at location 1; $c_1$: choice criterion for a response to location 1. 
({\bf B}) Performance in a 2-ADC task modeled with two binary choice (Yes/No) one-dimensional models, one at each potential target location. The subject employs independent choice criteria ($c_1, c_2$) at each location (red and blue vertical lines, respectively) while making a decision to respond in one of three ways: saccade to location 1, to location 2, or to withhold response (NoGo) (see text for details). 
({\bf C}) Multidimensional signal detection model for the 2-ADC task. The decision variable at each stimulus location ($\Psi_1, \Psi_2$) are represented along orthogonal axes in perceptual decision space. Decision variables are independently distributed. Black circle: schematic representation of a contour of the joint decision variable distribution for no stimulus at either location (noise distribution). Red circle: joint distribution of the decision variables for a stimulus at location 1. Blue circle: joint distribution of the decision variables for a stimulus at location 2. Linear decision boundaries (thick black lines) specify the domain of values of the decision variables for each potential response or choice. The integral of the decision variable distribution enclosed by each decision boundary represents the probability of the corresponding response: NoGo (Y=0, gray), Go response to location 1 (Y=1, red) or to location 2 (Y=2, blue). Marginal distributions of each decision variable are also shown alongside each axis.
}
\label{fig_model}
\end{figure}

\subsubsection*{2-ADC latent variable formulation}
\addcontentsline{toc}{subsubsection}{2-ADC latent variable formulation}

We formulate the 2-ADC model analytically building upon a recently developed latent variable formulation \cite{decarlo}. This formulation involves specifying a \textit{structural model} of the subject's perceptual sensitivity for detecting the presented stimulus, and a \textit{decision rule} that models the effect of choice bias on the subject's response. In the Discussion, we analyze the assumptions inherent in this formulation and discuss potential extensions. 

We denote the subject's response with the variable $Y$: $Y = i$ indicates that the subject chose to respond at location i (Go response) whereas $Y = 0$ indicates that the subject gave a NoGo response. Similarly, the presentation of stimulus at location $i$ is denoted by the variable $X_i$: $X_i = 1$ indicates that a stimulus presented at location $i$. We further stipulate that no more than one stimulus be presented on a given trial, a common practice in psychophysics tasks as well as in all tasks employed in this study. Thus, $\sum\limits_{k=1}^m X_k = 1$ (stimulus trial) or $0$ (catch trial). 

The structural model for 2-ADC task  independently distributed decision variables $\Psi_i$ for each of the two locations, and specifies how these distributions change for each stimulus condition:
\begin{eqnarray}
\Psi_1 & = & d_1 X_1 + \varepsilon_1 \nonumber \\ 
\Psi_2 & = & d_2 X_2 + \varepsilon_2 \label{eq:strmod}
\end{eqnarray}
\noindent

where $\Psi_i$ denotes the decision variable at location $i$, $\varepsilon_i$ is a random variable that represents the distribution of $\Psi_i$ when no stimulus is present, $d_i$ represents the change in the expected value of the decision variable when a stimulus is presented at location $i$ (vs. when no stimulus is present); in other words, $d_i = E(\Psi_i|X_i=1) - E(\Psi_i|X_i=0)$. Note that in this formulation each $\Psi_i$ can be considered a component of a bivariate random variable ($\mathbf{\Psi}$) represented in a two-dimensional, Cartesian, perceptual/decision space (Figure 2C).

The distribution of $\Psi_i$ when a stimulus is present at location $i$ ($X_i = 1$) is termed the ``signal'' distribution; its distribution when no stimulus is present ($X_i = 0$) is termed the ``noise'' distribution, and is identical with the distribution of $\varepsilon_i$. 

In line with conventional SDT, the 2-ADC structural model posits that a stimulus shifts the mean of the noise distribution (additively) without altering its variance or higher moments. Conventional SDT also assumes that the noise distribution is unit normal (zero-mean Gaussian with unit variance). If the noise distributions at the two locations are zero-mean Gaussian but have unequal variances, the 2-ADC structural model with unit normal noise distributions (\ref{eq:strmod}) can be readily recovered upon scaling each decision variable by the standard deviation of the appropriate noise distribution ($\sigma_i$). 

The terminology for $d_i$ deserves careful mention. $d_i$ measures the difference between the expected values of the signal and noise distributions, and is hence a measure of \textit{signal strength}. When measured in units of noise standard deviation ($\sigma_i$), $d_i$ is a measure of the overlap between signal and noise distributions and, hence, an \textit{index of discriminability} for these distributions. In a behavioral context, it is a measure of the subject's sensitivity for detecting the signal from a noisy background. For the purposes of this study, we will refer to $d_i$ as \textit{perceptual sensitivity} with the understanding that $d_i$ is measured in units of noise standard deviation. 

The SDT decision rule for a binary choice (Yes/No) task specifies that the subject reports a detection (``Yes'' response) if the decision variable exceeds a cutoff value, known as a choice criterion ($c$). The 2-ADC decision rule extends the one-dimensional SDT decision rule by specifying choice criteria for each location:
\vspace{-10pt}
\begin{eqnarray} \label{eq:decrule}
Y = 1, & \mathrm{if} & \Psi_1 > c_1 \ \cap \ \Psi_1 - c_1 > \Psi_2 - c_2 \nonumber \\
Y = 2, & \mathrm{if} & \Psi_2 > c_2 \ \cap \ \Psi_2 - c_2 > \Psi_1 - c_1 \\
Y = 0, & \mathrm{if} & \Psi_1 \leq c_1 \ \cap \ \Psi_2 \leq c_2 \nonumber
\end{eqnarray}

Thus, the subject makes a response at location $i$ when the decision variable $\Psi_i$ exceeds choice criterion $c_i$. If the $\Psi_i$-s exceed the choice criterion at both locations, then the subject responds to the location with the larger difference between decision variable and criterion values (larger $\Psi_i - c_i$). On the other hand, if $\Psi_i$-s fall below the choice criterion at every location, then the subject makes a NoGo response. 

The decision rule is depicted in Figure 2C (thick black lines). In a later section, we demonstrate how this rule can be derived from optimal decision theory (for the more general m-alternative case). Note that both sensitivities and criteria are measured in the same units: the assumption of unit normal noise is made with the understanding that both $d_i$-s and $c-i$-s are measured in units of noise standard deviation.   

These choice criteria $c_i$ constitute an SDT measure of bias. The relative value of the criteria between locations indicates the magnitude of the bias: a lower choice criterion at a location corresponds to a greater choice bias for that location.

In order to measure the contribution of bias to the observed responses, an analytical relationship must be formulated between the criteria, sensitivities and response probabilities. The structural model and decision rule permit establishing such a relationship. Here, we summarize the dependence of response probabilities on sensitivities and criteria; the detailed derivation is provided in the Methods.

The following system of equations constitute the 2-ADC model.
\begin{empheq}[box=\myyelbox]{align} \label{eq:sixindep3}
p(Y = i | X_i, X_j) &= \int_{c_i - d_i X_i}^{\infty}  F_j(e_i + d_i X_i - d_j X_j - (c_i - c_j)) \ f_i(e_i) \ de_i \\ 
p(Y = 0 | X_i, X_j) &=  F_i(c_i - d_i X_i) \ F_j(c_j - d_j X_j) & \nonumber \\
& \footnotesize{i,j \in \{1,2\}, i \neq j} \nonumber
\end{empheq}

where $p(Y = i | X_i, X_j)$ represents the conditional probability of a Go response to location $i$, for a particular stimulus contingency ($X_i, X_j$), and $p(Y = 0 | X_i, X_j)$ represents the conditional probability of a NoGo response for the given stimulus contingency in the 2-ADC task; $f_i$ and $F_i$ represent, respectively, the probability density function and the cumulative density function of the noise distribution $\varepsilon_i$ at location $i, i \in \{1, 2\}$.

\subsection*{Generalization to the multiple alternative detection (m-ADC) task}
\addcontentsline{toc}{subsection}{Generalization to the multiple alternative detection (m-ADC) task}

The multiple alternative (m-alternative) detection task is an extension of the 2-ADC task that permits more than two Go response alternatives along with the NoGo response alternative. We formulate the model next.

\subsubsection*{m-ADC latent variable formulation}
\addcontentsline{toc}{subsubsection}{m-ADC latent variable formulation}

The formulation of the m-ADC model is conceptually similar to that of the 2-ADC model. We estimate the criterion and sensitivity at each location based on a structural model and decision rule that are extensions of their 2-ADC counterparts.

\noindent
The m-ADC structural model is defined as follows:
\begin{align}
\Psi_i &= d_i X_i + \varepsilon_i & \varepsilon_i & \sim \mathcal{N}(0,1) & i \in \{1,2, \ldots m \} \label{madc:SM}
\end{align}

\noindent
The m-ADC decision rule is defined as follows: 
\begin{eqnarray}
Y& =& i \ \mathrm{if} \ \Psi_i > c_i \ \ \cap \nonumber \\
 &  & \ \ \ \ (\Psi_i - c_i) = \mathrm{max}(\Psi_1 - c_1, \Psi_2 - c_2, \ldots, \Psi_m - c_m) \nonumber \\ 
 & =& 0 \ \mathrm{if} \ \cap_k \ \Psi_k \leq c_k \label{madc:DR}
\end{eqnarray}

Thus, the subject gives a Go response to the location at which the decision variable exceeds the choice criterion at that location, and at which the difference between the decision variable and the corresponding choice criterion is maximal. If the decision variable does not exceed the choice criterion at any location, the subject gives a NoGo response.

As with the 2-ADC model, each $\Psi_i$ in the m-ADC model can be considered an independent component of a multivariate (random) decision variable ($\mathbf{\Psi}$) represented in a multi-dimensional perceptual space. In addition, the assumption of orthogonality (independence) among the perceptual dimensions implies that the covariance matrix of this decision variable is a diagonal matrix.

As before, the structural model and decision rule permit establishing the relationship between sensitivity, criteria and m-ADC response probabilities (derived in the Methods).

\begin{empheq}[box=\myyelbox]{align} 
p(Y = i | \mathbf{X}) &= \int_{c_i - d_i X_i}^{\infty} \prod_{k, k \neq i}  F_k(e_i + d_i X_i - d_k X_k - c_i + c_k) \ f_i(e_i) \ de_i \label{eq:maufc}
\end{empheq}

\noindent
where, as before $p(Y = i | \mathbf{X})$ represents the conditional probability of a Go response to location $i$, for a particular stimulus contingency ($\mathbf{X})$), and $p(Y = 0 | \mathbf{X})$ represents the conditional probability of a NoGo response for the given stimulus contingency in the m-ADC task; $f_i$ and $F_i$ represent, respectively, the probability density function and the cumulative density function of the noise distribution $\varepsilon_i$ at location $i, i \in \{1, \ldots, m\}$.

The system of equations (\ref{eq:maufc}) corresponds to $m^2+m$ independent observations from which the criteria and sensitivities at each location may be determined.

In the m-ADC task, $m^2+m$ independent probabilities are determined by just $2m$ parameters corresponding to the criterion and sensitivity at each of the $m$ locations. Thus, for the m-ADC case ($m>2$), there are far more equations than parameters compared to the 2-ADC case. Hence, there are more observations for internal tests of model validity.

\subsection*{Parameter estimation, uniqueness, and optimality}
\addcontentsline{toc}{subsection}{Parameter estimation, uniqueness and optimality}

In the two alternative forced choice (2-AFC) task, perceptual sensitivity and choice criteria are readily estimated analytically, as these quantities occur as linear terms of the argument of an invertible probit function \cite{greenswetsbk}. Moreover, the specification of a criterion (or cut-off value) in the 2-AFC model is Bayes optimal in terms of maximizing reward or the proportion of correct responses \cite{luce1963}.

On the other hand, the potential for multiple stimulus contingencies ($m>2$) and catch trials renders the m-ADC model multidimensional (Figure 2C) and raises several challenges. First, in this multidimensional SDT model, the system of equations (\ref{eq:maufc}) is not readily invertible analytically. Thus, given a set of experimentally observed m-ADC responses (e.g, contingency table, Table S1B), is it possible to solve the system of equations (\ref{eq:maufc}) to estimate the underlying perceptual sensitivity and choice criterion for each location? Second, having estimated these parameters, can one guarantee uniqueness, so that only one set of parameters is consistent with a given set of response probabilities? Finally, can one show that the specification of independent criteria at each location (linear, intersecting decision surfaces, Figure 2C) constitutes an optimal decision rule? We addressed these challenges with a combination of numerical and analytical approaches.

\subsubsection*{Estimating m-ADC model parameters}
\addcontentsline{toc}{subsubsection}{Estimating m-ADC model parameters}

\begin{figure}[!t]
\begin{center}
\includegraphics[width=4in]{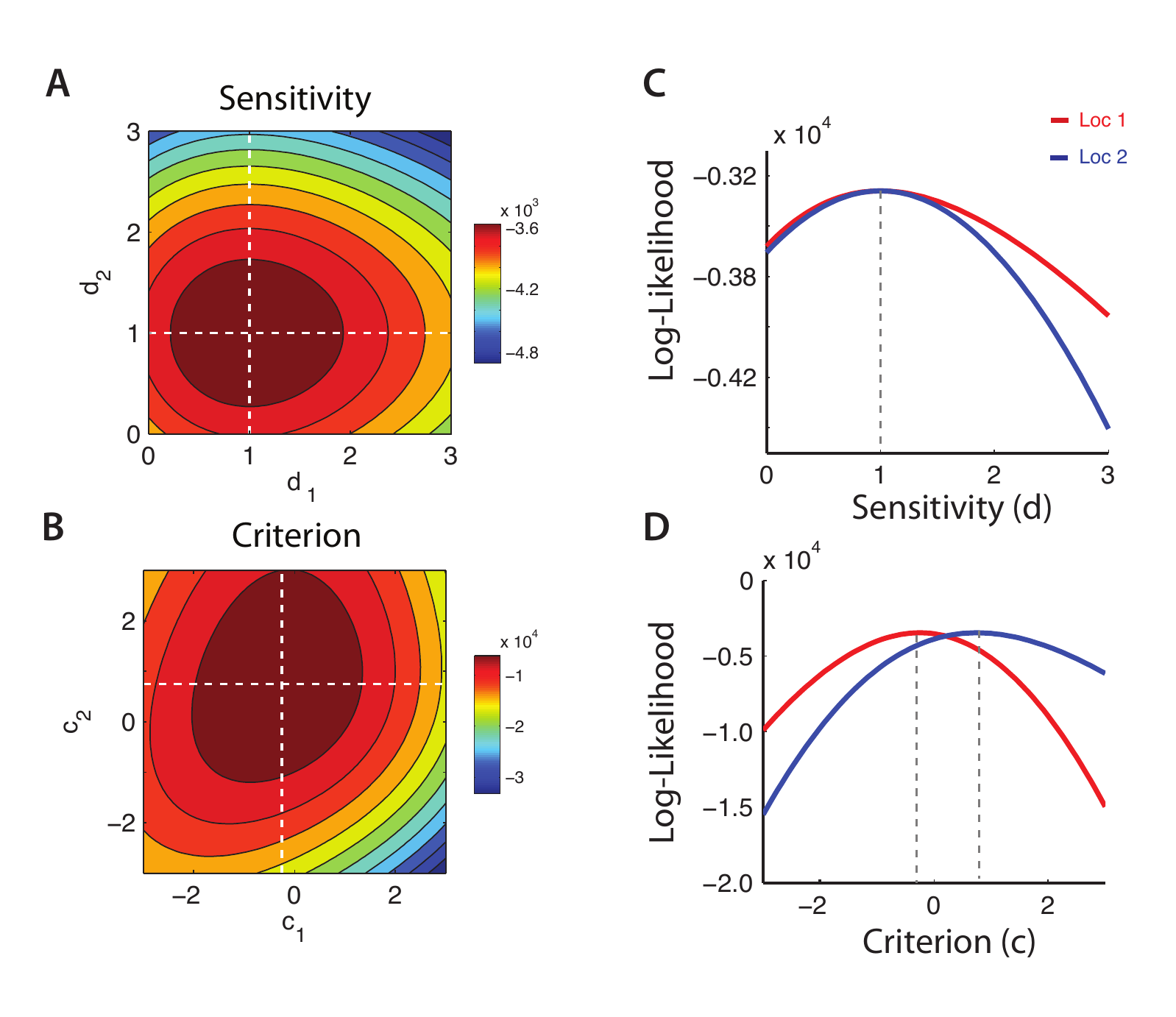}
\end{center}
\caption{
{\bf Likelihood landscape for the 2-ADC model.} 
({\bf A}) Contour plot of the 2-ADC multinomial log-likelihood as a function of the sensitivities ($d_1, d_2$) at the two locations. 
({\bf B}) Contour plot of the 2-ADC multinomial log-likelihood as a function of the criteria ($c_1, c_2$). The convexity of the function is apparent throughout the domain shown. 
({\bf C}) The variation of log-likelihood with sensitivity each location for fixed values of the other parameters (sensitivity at the other location and the two criteria, cross section through the dashed white lines of panels A-B).  Dashed gray lines: values of the parameters that maximize the log-likelihood function; red data: location 1; blue data: location 2.
({\bf D}) Same as panel C, but variation with the criterion at each location for fixed values of the other parameters (criterion at the other location, and the two sensitivities). 
}
\label{fig_errf}
\end{figure}

We employed numerical methods to estimate m-ADC model parameters, sensitivities and criteria, based on m-ADC response probabilities. We demonstrate parameter recovery with the 2-ADC model; the procedure can be readily extended to the m-ADC case.

The shape of the likelihood function is key to identifying the effectiveness of numerical approaches for parameter estimation, and for identifying the uniqueness of the underlying parameters. Responses to each alternative for each stimulus contingency ($n(Y=i|X_i, X_j)$) were assumed to follow a multinomial distribution.  We depict the four-dimensional likelihood function in a pair of two-dimensional subspaces by holding $c_i$-s constant and varying $d_i$-s or vice versa (Figure 3A-B, parameter values in Table S2A). In the domain of parameter values shown in the figure, the likelihood function appears to be convex (Figure 3A-D) indicating a single minimum corresponding to a unique set of underlying parameters.

As proof-of-concept of parameter recovery, simulated data were provided as input to the algorithm in lieu of experimental data. Simulated data were generated as follows: response probabilities were computed from (\ref{eq:sixindep3}) based on a prespecified set of criteria and sensitivities (Table S2A). We denote the response probabilities as $\mathcal{P}^r_s$. Based on these probabilities, response counts for each stimulus-response contingency were generated with random sampling. This procedure was repeated for 20 simulated runs with 100 trials for each of the two stimulus conditions, and 200 catch trials per run (a total of N=4000 trials in 20 runs). The resulting total response counts, $\mathcal{O}^r_s$ (Table S2B) were provided to numerical optimization algorithms for parameter recovery.

We employed two approaches: (i) maximum likelihood estimation with a line-search (ML-LS) algorithm or (ii) Bayesian estimation based on Markov Chain Monte Carlo approach with the Metropolis algorithm (MCMC, Methods). The ML-LS line-search algorithm is an efficient approach for maximum-likelihood estimation, but could converge onto a local extremum of the objective function. The MCMC algorithm, although comparatively slower, has a component of stochastic sampling (Methods), and hence, a better chance of finding global minima. In addition, errorbars computed with the MCMC approach provide independent confirmation of the parameter standard errors computed with the ML-LS algorithm.

Both the ML-LS and MCMC algorithms converged reliably onto identical values of the four parameters for various initial guesses (Table S2C). Figure 4A-B and 4C-D show the convergence of the ML-LS and MCMC algorithms, respectively, for various initial guesses of the four parameters  $\{d_i, c_i\}, i \in \{1,2\}$. The search trajectory in four-dimensional space is depicted as 2 two-dimensional trajectories, one for each pair of criterion and sensitivity parameters. The MCMC algorithm required an initial \textit{burn-in} period (about 500 iterations, Figure 4E), to converge to a stable parameter set; the $\chi^2$ error value reduced and the log-likelihood value increased systematically over successive (Figure 4F). The posterior distribution was generated with the parameter values from the last 1000 iterations, well after the burn-in period of the MCMC algorithm (Figure 4G, Methods). Error estimates of the parameters were also highly similar between the two estimation approaches (Table S2C).

\begin{figure}[!t]
\begin{center}
\includegraphics[width=6.5in]{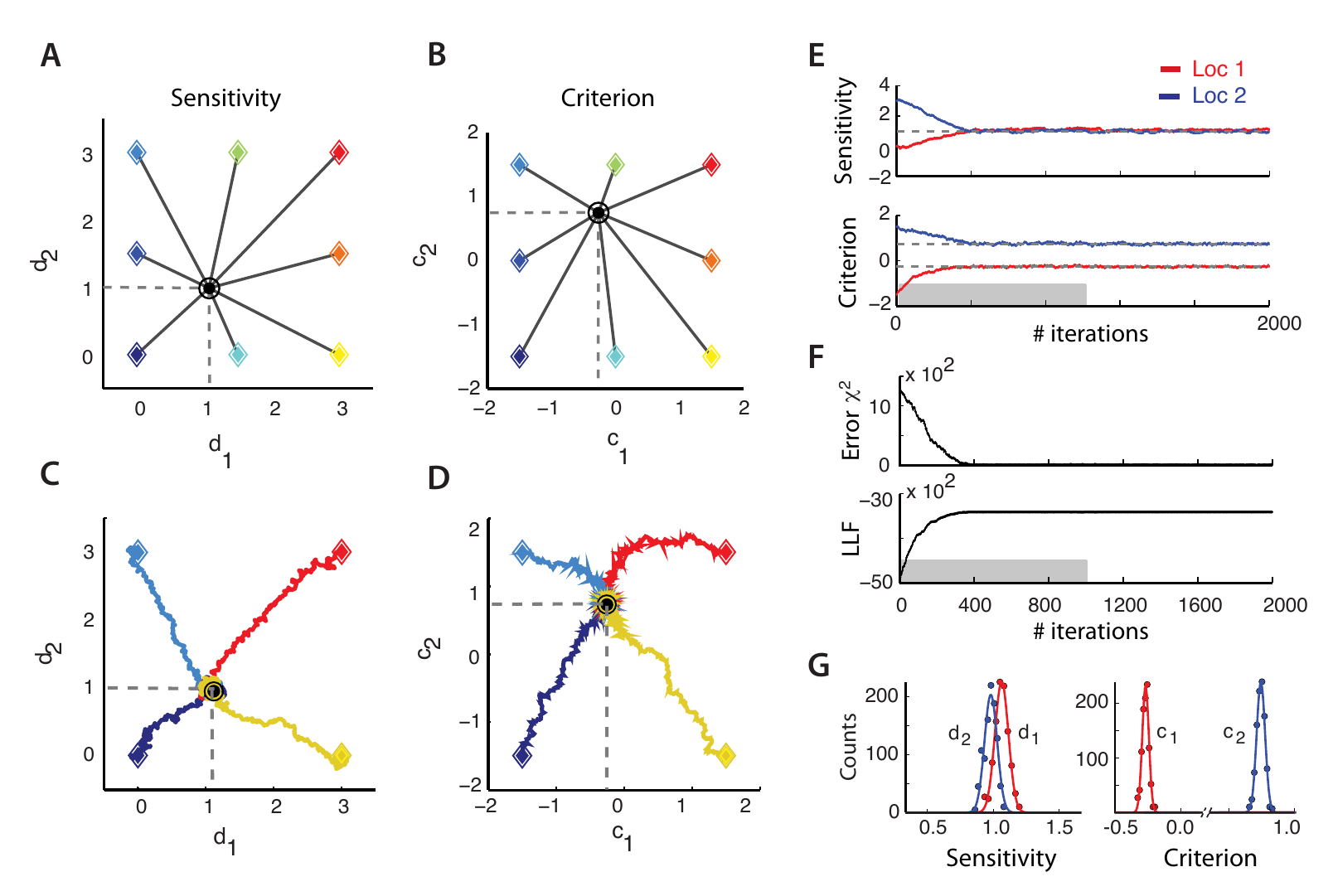}
\end{center}
\caption{
{\bf Estimating sensitivities and criteria from simulated 2-ADC responses.} 
({\bf A-B}) Maximum likelihood estimation of the perceptual sensitivity ({\bf A}) and choice criterion ({\bf B}) at each location from simulated response counts in the two-alternative detection task (Table S2B). Beginning with an initial guess for each parameter, the algorithm uses a line-search method to identify the sensitivities and criteria that maximize the likelihood of the simulated response counts. For various initial guesses (colored diamonds-s), the MLE algorithm converged reliably onto identical values of sensitivity and criterion values at each location (black circles/dashed gray lines).
({\bf C-D}) Markov-chain Monte-Carlo (MCMC, Metropolis sampling) algorithm for estimating perceptual sensitivity ({\bf C}) and choice criterion ({\bf D}) at each location from simulated 2-ADC responses (same as in panels A-B). For various initial guesses (colored diamonds-s), the Markov chain converged reliably to identical values of sensitivity and criterion at each location (black circles). Colored lines: Markov chains during MCMC runs for different initial guesses. 
({\bf E}) Evolution of the values of sensitivity (upper panel) or criterion (lower panel) at each location during a particular MCMC run (magenta data in panels C-D) for location 1 (red) or location 2 (blue). Gray bar: burn-in period (1000 iterations).
({\bf F}) The chi-squared error function (upper panel) decreased steadily, and the log-likelihood increased (lower panel) over successive iterations of the MCMC run. 
({\bf G}) Stationary (posterior) distributions (circles) of the sensitivity (left panel) and criterion (right panel) values at each location for the MCMC run (panel E). These distributions were used to construct standard errors and 95\% credible intervals for the parameters (Table S2C). Red data: location 1; blue data: location 2. Lines: Gaussian fits to each distribution.
}
\label{fig_mlemcmc}
\end{figure}

Thus, the 2-ADC model (\ref{eq:sixindep3}) can be readily solved numerically to estimate independently the sensitivities and criteria from observed response counts in the contingency table. In addition, the reliable convergence to a unique set of parameters (Figure 4A-D) indicates a single maximum of the likelihood function (Figure 3).


\subsubsection*{Uniqueness of m-ADC model parameters}
\addcontentsline{toc}{subsubsection}{Uniqueness of m-ADC model parameters}

In the previous section, we demonstrated that criterion and sensitivity at each location may be estimated, with numerical approaches, from the response probabilities. The ability to reliably recover an identical set of parameters with both the ML-LS and MCMC algorithms for various initial values of parameter guesses suggests that the solution to the system of equations (\ref{eq:maufc}) is unique. However, the equations (\ref{eq:maufc}) are non-linear integral equations, and we must entertain the possibility that multiple parameter configurations may be consistent with a given set of response probabilities. Is the solution to system (\ref{eq:maufc}), corresponding to the estimated set of criteria and sensitivities, unique?

We addressed this question analytically. We illustrate the approach with the 2-ADC model, and then extend this to the m-ADC model.

Consider the effect of varying sensitivities and criteria on response probabilities in the 2-ADC model. Each of the nine response probabilities in the model (\ref{eq:sixindep3}) is a function of the four parameters: the criterion and sensitivity at each of the two locations. Thus, each response probability constitutes a surface in four-dimensional parameter space ($\{d_i, c_i\}, i \in \{1,2\}$). To facilitate representation, we examined a pair of two-dimensional subspaces by varying the criteria holding the sensitivities constant (parameter values in Table S2A), and vice versa. In line with conventional SDT, noise was assumed to be normally distributed with zero-mean and unit variance. 

The task specification requires that no more than one stimulus be presented on a given trial (see Section 1). This permits us to employ the following notational short-hand for the response probabilities: $p(Y=i| X_j=1) = p^i_j$, where the superscript denotes the response location and the subscript denote the stimulus location.

Figure 5A illustrates the effect of varying criterion $c_i$ at each location ($i \in \{1,2\}$), on the response probabilities at a particular location say, location 1. The following general trends are apparent from the figure: A higher choice criterion at a location $i$ (lower bias toward location $i$) reduces the probability of response at that location ($p^i_k$) and enhances the probability of response at the opposite location ($p^j_k, j \neq i$), regardless of where the stimulus is presented ($i,j,k \in \{1,2\}$). 

Also apparent is the effect of sensitivity ($d_i$) on response probabilities: Greater sensitivity to a stimulus at a location enhances the hit-rate at that location (Figure 5B, red), and reduces the probability of a false alarm (incorrect response) at the opposite location (Figure 5B, blue). 

The monotonic dependence of the response probabilities on each criterion and sensitivity is analytically formulated in Lemmas 1-3 (see below) and demonstrated in the Appendices A.1-A.3. These monotonic relationships permitted us to establish that a given set of response probabilities uniquely determines the underlying sensitivities and criteria in the 2-ADC model.

\paragraph*{I. Uniqueness of 2-ADC sensitivities and criteria \\}
\addcontentsline{toc}{paragraph}{I. Uniqueness of 2-ADC sensitivities and criteria}

In this section, we demonstrate that the set of parameters $\{d_i, c_i\}, i \in \{1,2\}$ constitute a unique solution of the system of equations (\ref{eq:sixindep3}), i.e. they uniquely determine the 2-ADC response probabilities. In the previous section, we conjectured that the 2-ADC likelihood function is convex over the entire domain of sensitivity and criteria (for a Gaussian noise distribution) indicating a unique, global minimum: a potential approach is, then, to demonstrate analytically the convexity of the likelihood function (Figure 3). Such a demonstration is necessarily tied to a specific choice of likelihood function and noise distribution ($f$). Here, we employ a different approach, based on analytical reasoning and mathematical induction, that is free of these assumptions. 

The proof proceeds in two steps. In the first step, we demonstrate the uniqueness of the choice criteria. In the second step, we build upon the previous result to demonstrate the uniqueness of the perceptual sensitivities.

\paragraph{I-A. Uniqueness of the 2-ADC criteria}
First we consider the system of response probabilities when no stimulus was presented, i.e. false-alarm rates at each location during catch trials (as mentioned before, we use $p^i_0$ as a notational shorthand for $p(Y=i|X_i=0, X_j=0)$): 
\begin{eqnarray} 
p^1_0 & = & \int_{c_1}^{\infty}  F_2(e_1 - c_1 + c_2) \ f_1(e_1) \ de_1 \nonumber \\
p^2_0 & = & \int_{c_2}^{\infty}  F_1(e_2 - c_2 + c_1) \ f_2(e_2) \ de_2  \label{eq:pi0a}
\end{eqnarray}

We demonstrate that if a solution $(c_1, c_2)$ of the system exists, then the solution is unique. The analytical proof rests on the following lemmas: 
\begin{quote}
\textbf{Lemma 1} \ \ $p^i_0(c_i, c_j)$ is a monotonically decreasing function of $c_i$ and a monotonically increasing function of $c_j$, $i,j \in \{1,2\}, i \neq j$. 

\textbf{Lemma 2} \ \ $p^0_0$ is a monotonically increasing function of both $c_1$ and $c_2$. Specifically, \\ $p^0_0 = F_1(c_1) \ F_2(c_2)$. 
\end{quote}

\noindent
Simply put, these lemmas assert that response probabilities vary monotonically as a function of choice criteria. The proof of these lemmas is provided in Appendices A.1-A.2 (Supporting Information). The proof assumes no specific form for the functions $f_1$ and $f_2$; only that they are continuous and supported over the entire domain of integration. Upon rearrangement of the identity in Lemma 2: 
\begin{align} \label{eq:cmon}
F_i(c_i)& = p^0_0 / F_j(c_j)& i,j \in \{1,2\}, i \neq j
\end{align}

\noindent
The sequence of arguments for the proof follows: 
\begin{enumerate}[(i)]
 \item Let $(c_1, c_2)$ be a particular solution for a given value $\mathcal{P}^i_0$ of $p^i_0$. Assume that there exists another solution $(c_1', c_2')$, such that at least one $c_i'$ is distinct from $c_i$. 
 \item Without loss of generality, let $c_1 > c_1'$. 
 \item From Lemma 1, \\ $c_1 > c_1' \Rightarrow p^1_0(c_1, c_2) = \mathcal{P}^1_0 < p^1_0(c_1', c_2)$. Similarly, \\ $c_1 > c_1' \Rightarrow p^2_0(c_1, c_2) = \mathcal{P}^2_0 > p^2_0(c_1', c_2)$. 
 \item Hence, it follows that $c_2' > c_2$ for constant $\mathcal{P}^i_0$. In other words, if one choice criterion \textit{increases}, the other must also \textit{increase} to keep $p^1_0$ constant.
 \item Being cumulative distribution functions, $F_i$-s are monotonic functions of their arguments. Thus, $F(c_1) > F(c_1') \Leftrightarrow c_1 > c_1'$. 
 \item From Lemma 2, (equation (\ref{eq:cmon})), and point (v) above: $c_1 > c_1' \Rightarrow c_2' < c_2$ for constant $\mathcal{P}^0_0$. In other words, if one choice criterion \textit{increases}, the other must \textit{decrease} to keep $p^0_0$ constant.
\end{enumerate}
The proof follows by contradiction.

\paragraph*{I-B. Uniqueness of the 2-ADC sensitivities}
Next, we demonstrate that a unique solution $(c_1, c_2)$ of the system of equations (\ref{eq:pi0a}) corresponding to response probabilities in the no-stimulus condition implies a unique solution $(d_1, d_2)$ of the system of equations corresponding to response probabilities when a stimulus was presented. 
 
For a stimulus presented at location $i$, response probabilities at location $i$ are given by (refer equations (\ref{eq:sixindep1}) and (\ref{eq:sixindep2})):
\begin{eqnarray} 
p^1_1 & = & \int_{c_1 - d_1}^{\infty}  F_2(e_1 + d_1 - c_1 + c_2) \ f_1(e_1) \ de_1 \\
p^2_2 & = & \int_{c_2 - d_2}^{\infty}  F_1(e_2 + d_2 - c_2 + c_1) \ f_2(e_2) \ de_2 
\end{eqnarray}

The proof rests on the following lemma, which is proved in Appendix A.3 (Supporting Information): 
\begin{quote}  
\textbf{Lemma 3} \ \ $p^i_j(d_i, d_j)$ is a strictly monotonic function of its arguments $(d_i, d_j),  i,j \in \{1,2\}, i \neq j$. 
\end{quote} 

The sequence of arguments for the proof follows:
 
\begin{enumerate}[(i)]
\item Based on the previous section, we have already established a unique pair of criteria $(c_1, c_2)$ that satisfies two out of the six equations of system (\ref{eq:sixindep3}). Thus, the criteria are fixed based on the response probabilities during catch trials (false alarms and correct rejections).
\item Given a particular $(c_1, c_2)$, each of the probabilities, $p^1_1$ and $p^2_2$, in the above system of equations is only a function of its respective $d_i, i \in \{1,2\}$. 
\item By Lemma 3, $p^i_i$ is a strictly monotonic function of its respective $d_i$.
\item Strict monotonicity of each function implies that the transformation from the sensitivity to the response probability is invertible. Thus, if $d_i$ uniquely determines $p^i_i$ then, $p^i_i$ uniquely determines a $d_i$. In other words, there is a one-to-one mapping between the $p^i_i$-s and the $d_i$-s.
\item Hence, there is a unique solution $d_i$ that satisfies each of these equations.
\end{enumerate}

We have shown that given a set of choice criteria, there is a unique set of sensitivity parameters that satisfy (\ref{eq:sixindep3}). In the previous section, we showed that the choice criteria are also unique. Thus, the set of parameters $\{d_i, c_i\}, i \in \{1,2\}$ constitute a unique solution of the system of equations (\ref{eq:sixindep3}). This completes the proof. 

A geometric intuition for the proof may be obtained from Figure 2C by attempting to vary the criteria $c_1$ and $c_2$ and observing the effect on the response probabilities to each alternative $Y = 0,1,2$. These response probabilities correspond to the area under the joint probability distribution of $\Psi_1$ and $\Psi_2$ in each of the three decision regions. 

To illustrate this intuition, consider the response probabilities in the 2-ADC parameter space of criteria during catch trials (Figure 5C, $d_1 = d_2 = 0$). The sets of all possible pairs of choice criteria that could determine the probability of each type of response during catch trials (locus of variation of $c_1$ and $c_2$ for specific values of $p^i_0$) are shown in Figures 5C-D (colored contours; $i=1$, red; $i=2$, blue; $i=0$; green). Note that the three contours intersect at exactly one point in the $c_1-c_2$ plane (Figure 5D), indicating that exactly one pair of criteria is consistent with these response probabilities. Given these criteria values, the monotonic (one-to-one) relationship between perceptual sensitivity and response probability (Figure 5B) demonstrates that a specific value of the response probability is consistent with one, and only one, value of sensitivity at a location. The analytical proof presented above simply formalized this graphical demonstration.

\begin{figure}[!t]
\begin{center}
\includegraphics[width=5.5in]{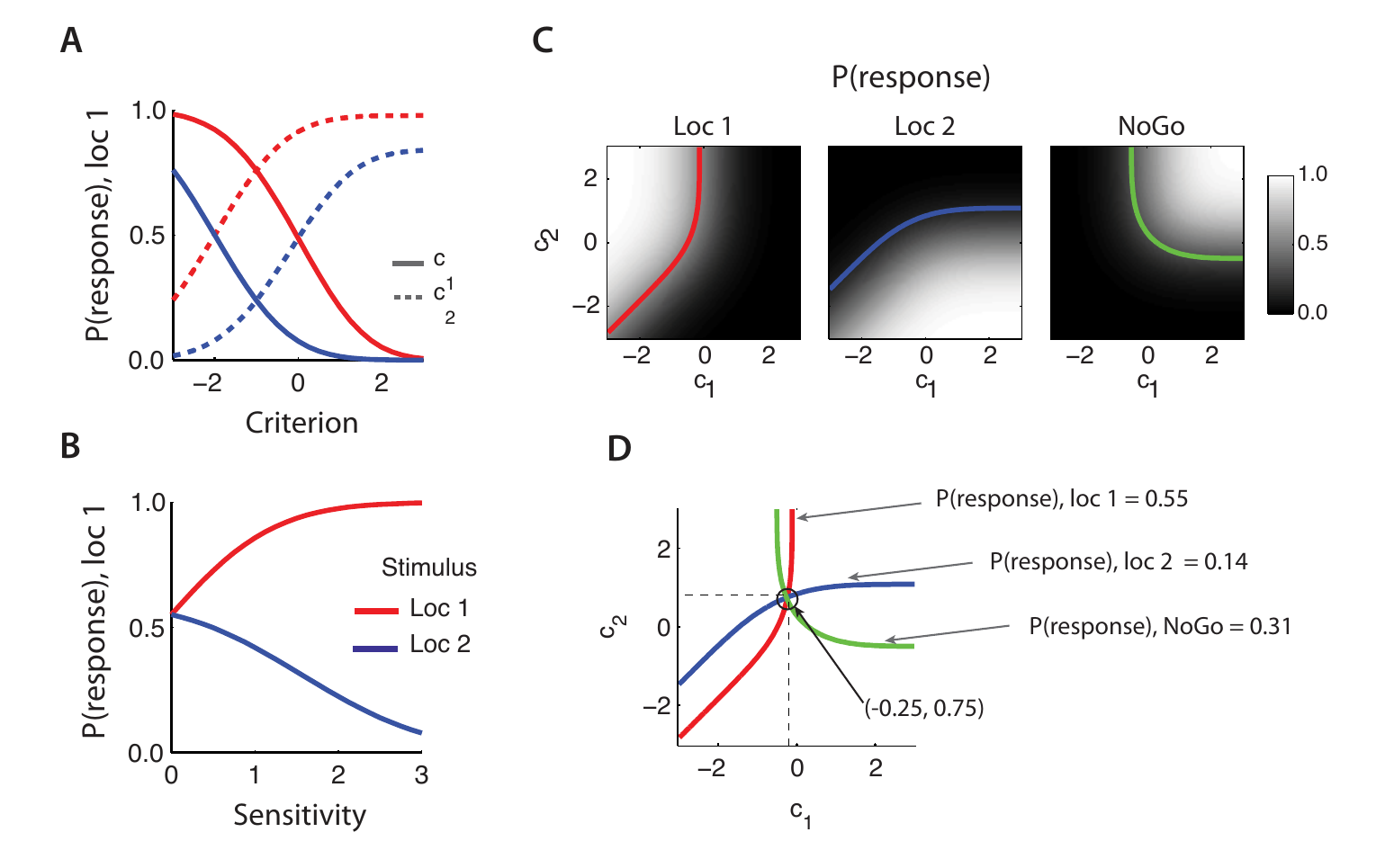}
\end{center}
\caption{
{\bf Uniqueness of 2-ADC model parameters.}  
({\bf A}) Variation of the probability of response at location 1 with the criterion at each location (for constant sensitivities, Table S2A). The probability of response to location 1, for a stimulus presented at location 1, decreases monotonically with an increasing choice criterion ($c_1$) at location 1 (solid red line) and increases monotonically with an increasing choice criterion ($c_2$) at location 2 (dashed red line). The same monotonic trends are observed when a stimulus is presented at location 2 (blue curves). 
({\bf B}) Variation of the probability of response at location 1 with the sensitivity at each location (for constant criteria, Table S2A). The probability of response to location 1 increases monotonically with increasing sensitivity ($d_1$) to a stimulus at location 1 (red), and decreases monotonically with increasing sensitivity ($d_2$) to a stimulus at location 2 (blue). 
({\bf C}) Probability of response (during catch trials) to a stimulus at location 1 (left), location 2 (middle), or NoGo (right) as function of the choice criterion at each location. Colored lines: The contour traversing all possible pairs of criteria consistent with a specific value of each response probability; red: probability of a Go response to location 1; blue: probability of a Go response to location 2; green: probability of a NoGo response. 
({\bf D}) The three contours (red, blue, green) intersect at a single point indicating that exactly one set of criteria is consistent with a given set of response probabilities. Arrows: Specific values of NoGo and Go response probabilities at each location and the unique pair of criteria that is consistent with this specific set of response probabilities.
}
\label{fig_model}
\end{figure}

Note that the complementary probabilities, viz., response probabilities at location a when a stimulus was presented at the opposite location, corresponding to the following pair of equations:
\begin{eqnarray}
p^1_2 & = & \int_{c_1}^{\infty}  F_2(e_1 - d_2 - c_1 + c_2) \ f_1(e_1) \ de_1 \\
p^2_1 & = & \int_{c_2}^{\infty}  F_1(e_2 - d_1 - c_2 + c_1) \ f_2(e_2) \ de_2
\end{eqnarray}
 
\noindent 
have not been used to demonstrate parameter uniqueness. Conditioned on the 2-ADC model with four parameters ($d_1, d_2, c_1, c_2$), these two probabilities are not free to vary (system (\ref{eq:sixindep3}) is overdeterminate). If the solution $\{d_i, c_i\}$ does not satisfy these equations, we must consider the possibility that no solution to system (\ref{eq:sixindep3}) exists. In this case, more elaborate models may be necessary to fit the data, and these are discussed below. What we have shown in this section is that if a solution set, $\{d_i, c_i\}$, of parameters exists, then it is unique under this model.


\paragraph*{II. Uniqueness of m-ADC sensitivities and criteria \\}
\addcontentsline{toc}{paragraph}{II. Uniqueness of m-ADC sensitivities and criteria}

In the previous section, we showed that criteria and sensitivities are uniquely determined by response probabilities in the 2-ADC model. In this section we extend this result to demonstrate that response probabilities in the m-ADC model uniquely determine the $2m$ parameters, corresponding to the criterion and sensitivity at each of the $m$ locations, for any $m \ge 2$. 

The demonstration proceeds in two steps. First we demonstrate the following with mathematical induction: if the m-ADC model for a task with $m$ response alternatives has a unique solution set of criteria, then so does the model for a task with $(m+1)$ alternatives. Next, we utilize monotonicity to show that perceptual sensitivities are also uniquely determined for the m-ADC task. 

\paragraph*{II-A. Uniqueness of the m-ADC criteria}
We consider the probability of responses when during catch trials. This is given by:
\begin{align}
p(Y = i | \ \lVert\mathbf{X}\rVert_{1} = 0) &= \int_{c_i}^{\infty} \prod_{k, k \neq i}  F_k(e_i - c_i + c_k) \ f_i(e_i) \ de_i  \nonumber \\
& i \in \{0, \ldots, m\} \label{eq:madc_cth}
\end{align}

\begin{quote}
\textbf{Statement} \ \ Given a set of response probabilities $\mathcal{P}^i_0$ for an m-alternative task, if the system of equations (\ref{eq:madc_cth}) has a solution given by the ordered set of criteria $C = \{c_i: i \in \{0, \ldots, m\}\}$, then this solution is unique.
\vspace{12pt}

\textbf{Basis} \ \ The set of parameters $C = \{c_1, c_2\}$, forms a unique solution of the system  (\ref{eq:madc_cth}) for a 2-alternative task (m=2).
\vspace{12pt}

\textbf{Inductive step} \ \ Given a set of response probabilities $p^i_0=\mathcal{Q}^i$ for an \textit{m-alternative} task, let the set of criteria $C_m = \{c_i: i \in \{0, \ldots, m\}\}$ be a unique solution of the system (\ref{eq:madc_cth}). Then, for a given set of response probabilities $p^i_0=\mathcal{P}^i_0$ for an \textit{m+1-alternative} task, a set of criteria $C_{m+1} = \{c_j: j \in \{0, \ldots, m+1\}\}$ that satisfies system equations (\ref{eq:madc_cth}) is also a unique solution.
\end{quote}

\subparagraph*{Proof of basis} 
In a previous section we proved that the criteria are uniquely determined by the 2-ADC response probabilities during catch trials (see Section 3.3-I). This constitutes the proof of the basis for $m = 2$.

\subparagraph*{Proof of inductive step}
The inductive step is proved, as before, in two stages:

The proof rests on the following lemmas, which are proved in Appendices A.4-A.6 (Supporting Information). 
\begin{quote}
\textbf{Lemma 4} \ \ Given a set of response probabilities $p^r_0=\mathcal{P}^r_0, r \in \{0, \ldots, m+1\}$, and any solution set $C = \{c_j: j \in \{0, \ldots, m+1\}\}$ comprising ordered sets of choice criteria satisfying the system (\ref{eq:madc_cth}). There is a one-to-one mapping between any choice criterion $c_i$ and its complement set $C'_i = \{c_j: j \in \{0, \ldots, m+1\}, j \neq i\}$.

\textbf{Lemma 5} \ \ Given a set of response probabilities $\mathcal{P}^i_0, i \in {1, \ldots, m+1}$ and the set of all solution sets $\{C^k = \{{c_j}^k: j \in {0, \ldots, m+1}\}\}$ comprising ordered sets of choice criteria satisfying the system (\ref{eq:madc_cth}). For any two solution sets $C^1$ and $C^2$ every pair of corresponding elements $({c_j}^1, {c_j}^2)$ obeys the same order relation i.e. if any ${c_i}^1 \gtrless {c_i}^2$ then every ${c_j}^1 \gtrless {c_j}^2, i,j \in \{0, \ldots, m+1\}, i \neq j$.

\textbf{Lemma 6} \ \ Given a set of response probabilities $ \mathcal{P}^0_0 $ and the set of all solution sets $\{C^k = \{{c_j}^k: j \in {0, \ldots, m+1}\}\}$ comprising ordered sets of choice criteria satisfying the system (\ref{eq:madc_cth}). For any two solution sets $C^1$ and $C^2$ at least one pair of corresponding elements $({c_j}^1, {c_j}^2)$ differs in its order relation i.e. if any ${c_i}^1 \gtrless {c_i}^2$ then at least one ${c_j}^1 \lessgtr {c_j}^2, i,j \in \{0, \ldots, m+1\}, i \neq j$.
\end{quote}

Simply put, Lemma 4 states that given set of false alarm and correct rejection rates, fixing one choice criterion determines all of the other choice criteria. The proof of Lemma 4 utilizes the induction hypothesis (see Appendix A.4). Lemma 5 states that if the choice criterion to one location were to increase (decrease), the choice criterion at every location has to also increase (decrease) to maintain the false alarm rate unchanged at each location. Lemma 6 states that if the choice criterion to one location were to increase (decrease), the choice criterion at least at one location has to decrease (increase) to maintain the correct rejection rate unchanged.

The sequence of arguments for the proof proceeds as follows:
\begin{enumerate}[(i)]
 \item Let $C = \{c_j: j \in \{0, \ldots, m+1\}\}$ correspond to a solution of the system (\ref{eq:madc_cth}) for a specific value of $p^i_0=\mathcal{P}^i_0$. Let $C' = \{c_j': j \in \{0, \ldots, m+1\}\}$ be a different, unique solution for the same $\mathcal{P}^i_0$. 
 \item By Lemma 4, $c_j \neq c_j' \ \forall \  j$. Without loss of generality, let $c_i > c_i'$.
 \item By Lemma 5, if $c_i > c_i'$, then $c_j > c_j' \ \forall \  j, j \neq i$.
 \item By Lemma 6, if $c_i > c_i'$, then at least one $c_j < c_j'$ for some $j \neq i$.
\end{enumerate}

The proof follows by contradiction. Thus, there is a unique set of criteria $C = \{c_j: j \in \{0, \ldots, m+1\}\}$, which satisfies (\ref{eq:madc_cth}) for $p(Y = i | X_0=1)=\mathcal{P}^i_0$. In other words, specifying the response probabilities during catch trials (false alarm rates) at each of the $m$ locations uniquely determines the choice criteria.

\paragraph*{II-B. Uniqueness of the m-ADC sensitivities}
The proof rests on the following lemma (proved in Appendix A.7, Supporting Information):
\begin{quote}
\textbf{Lemma 7} \ \ The response probability $p^i_j(d_k)$ is a strictly monotonic function of each $d_k$.
\end{quote}

The sequence of arguments proceeds as follows:
\begin{enumerate}[(i)]
\item By the task specification, no more than one stimulus is presented on a given trial. Thus, for a fixed set of criteria $C$, the response probabilities $p^i_i$ of (\ref{eq:maufc}) are simply a function of their respective perceptual sensitivities $d_i$. 
\item From Lemma 7, the response probability $p^i_i(d_i)$ is a strictly monotonic function of its respective $d_i, i \in \{1, \ldots, M\}$. 
\item Strict monotonicity implies invertibility and demonstrates a one-to-one mapping between the $p^i_i$-s and the $d_i$-s.
\item Hence, there is a unique solution $d_i$ determined by each of the m probabilities $p^i_i, j \in \{1, \ldots, M\}$.
\end{enumerate}

This completes the proof. Note that the same arguments could be made with other sets of probabilities, such as the false-alarm rates, $p^j_i$, for reporting a stimulus at location $j$ when a stimulus was presented at location $i$, which are also monotonic functions of $d_i$ (Appendix A.7).

\subsubsection*{Optimality of m-ADC decision surfaces}
\addcontentsline{toc}{subsubsection}{Optimality of m-ADC decision surfaces}

We have demonstrated that the m-ADC model equations are solvable, and that the solution is unique. But does the model have the potential to explain real-world behaviors that typically involve strategies aimed at maximizing benefit (or reward) or minimizing cost (or punishment)? 

Here, we show that for the m-ADC formulation (\ref{madc:SM}), the m-ADC decision boundaries (\ref{madc:DR}) belong to a family of optimal decision surfaces for maximizing average utility or minimizing average risk; the utility (or risk) is defined as the benefit (or cost) associated with choosing a particular response when a particular stimulus occurs, and is assumed to be uniquely specified for each stimulus-response contingency. When the cost of making an erroneous response is the same for all stimulus contingencies, optimal decision surfaces for maximizing average utility (or minimizing average risk) for additive signals and noise are the isosurfaces of the generalized likelihood (or constant log-likelihood) ratio \cite{middleton}. 

For the m-ADC task, the decision variable (signal and noise) distributions (\ref{madc:SM}) can be rewritten in terms of a multivariate (m-dimensional) Gaussian random variable with a diagonal covariance matrix. The equation of such a multivariate Gaussian variable $\mathbf{W}$, with mean $\mathbf{W^0}=[w_1^0 w_2^0 \ldots w_m^0]$ and a diagonal covariance matrix $\mathbf{C}$ ($\mathbf{C}_{kk} = \sigma_k, \mathbf{C}_{kl} = 0, \ \ k,l \in \{1, ..., m\}, k \neq l$) is:

\begin{equation}
N(\mathbf{W}; \mathbf{W_0, C}) = A e^{-\sum\limits_{k=1}^m \ \frac{(w_k-w_k^0)^2}{2 \sigma_k^2}} 
\end{equation}

where $A = 1/\prod_{k=1}^m \sqrt{2 \pi} \sigma_k$ is a normalization constant in order for $N$ to be a probability density. 

The m-ADC model assumes that the noise distribution has zero mean ($\mathbf{W^0} = \mathbf{0}$). Thus, when no stimulus is presented ($\lvert \mathbf{X} \rvert_1 = 0$ or $X_i = 0 \forall i$), the noise distribution is given by:
\begin{equation}
N(\mathbf{W}| X_i=0 \ \ \forall i) = A e^{-\sum\limits_{k=1}^m \ \frac{w_k^2}{2 \sigma_k^2}} 
\end{equation}

A stimulus at location $j$ displaces the noise distribution along its perceptual dimension by $q_j$. Thus, the signal distribution in this case is given by: 
 
\begin{equation}
N(\mathbf{W}|  X_j=1, X_i=0 \ \ \forall i \neq j) = A e^{- \frac{(w_j-q_j)^2}{2 \sigma_j^2} - \sum\limits_{k=1, k \neq j}^m \ \frac{w_k^2}{2 \sigma_k^2} } 
\end{equation}

For the decision between the stimulus at location $j$ versus no stimulus, the log-likelihood ratio is given by: 
\begin{align}
\mathcal{L}_{j0}(\mathbf{W}) &= \log \frac{N(\mathbf{W}|  X_j=1, X_i=0 \ \ \forall i \neq j)}{N(\mathbf{W}| X_i=0 \ \ \forall i) } \\
&= \frac{-(w_j-q_j)^2}{2 \sigma_j^2} + \frac{w_j^2}{2 \sigma_j^2}
\end{align}

which after some rearrangement reduces to: 
\begin{align}
\mathcal{L}_{j0}(\mathbf{W}) &= \frac{2 w_j q_j - q_j^2}{2 \sigma_j^2} 
\end{align}

We define normalized quantities, $\Psi_j = w_j/\sigma_j$ and $d_j = q_j/\sigma_j$, measured in units of noise standard deviation along dimension $j$. Thus,
\begin{align}
\mathcal{L}_{j0}(\mathbf{W}) &= \Psi_j d_j - d_j^2/2
\end{align}

The generalized likelihood ratio is obtained by multiplying the ratio of the prior probabilities of stimulus $j$ and no stimulus ($p_j/p_0 = p(X_j=1, X_i=0 \ \ \forall i \neq j)/p(X_i=0 \ \ \forall i)$) with the likelihood ratio. Thus, the generalized log-likelihood ratio is obtained by adding $\log (p_j/p_0)$ to the likelihood ratio.
\begin{align}
\Lambda_{j0}(\mathbf{W}) &= \mathcal{L}_{j0}(\mathbf{W}) \ + \log \frac{p_j}{p_0} \\
& =  \Psi_j d_j - d_j^2/2 + \log \frac{p_j}{p_0} 
\end{align}

Thus, optimal decision surfaces for distinguishing a stimulus at location $j$ from no stimulus (isosurfaces of constant $\Lambda_{j0}$) are hyperplanes of constant $\Psi_j$. Thus, the specification of a cutoff criterion at $\Psi_j = c_j$ corresponds to the subject making a decision by selecting one from among the family of optimal decision surfaces. The precise choice of $c_j$ would depend on the cost/utility of choosing each alternative, the prior probability ratio ($p_j/p_0$) as well as the sensitivity along that dimension ($d_j$).  

Let us now consider the decision between a stimulus at location $i$ versus a stimulus at location $j$. For this case, the log-likelihood ratio is given by: 
\begin{align}
\mathcal{L}_{ij}(\mathbf{W}) &= \log \frac{N(\mathbf{W}|  X_i=1, X_k=0 \ \ \forall k \neq i)}{N(\mathbf{W}| X_j=1, X_k=0 \ \ \forall k \neq j) } \\
&= \frac{-(w_i-q_i)^2}{2 \sigma_i^2} + \frac{w_i^2}{2 \sigma_i^2} + \frac{(w_j-q_j)^2}{2 \sigma_j^2} - \frac{w_j^2}{2 \sigma_j^2}
\end{align}

which after some rearrangement reduces to: 
\begin{align}
\mathcal{L}_{ij}(\mathbf{W}) &= \frac{(2 w_i q_i - q_i^2)}{2 \sigma_i^2} - \frac{(2 w_j q_j - q_j^2)}{2 \sigma_j^2} 
\end{align}

As before, introducing normalized quantities $\Psi_i = w_i/\sigma_i, d_i = q_i/\sigma_i, \Psi_j = w_j/\sigma_j, d_j = q_j/\sigma_j$, measured in units of noise standard deviation along the respective dimensions ($i$ or $j$):
\begin{align}
\mathcal{L}_{ij}(\mathbf{W}) &= \Psi_i d_i - d_i^2/2 - \Psi_j d_j + d_j^2/2
\end{align}

And the generalized log-likelihood ratio is, as before: 
\begin{align}
\Lambda_{ij}(\mathbf{W}) & =  \Psi_i d_i - \frac{d_i^2}{2} - \Psi_j d_j + \frac{d_j^2}{2} + \log \frac{p_i}{p_j} 
\end{align}

Optimal decision surfaces for distinguishing a stimulus at location $i$ from a stimulus at location $j$ (isosurfaces of constant $\Lambda_{ij}$) are hyperplanes obeying the following relation: $\Psi_i d_i - \Psi_j d_j = \mathcal{C}$, where $\mathcal{C}$ is a constant that incorporates the benefit (or cost) associated with correctly identifying stimulus contingencies $i$ and $j$. Specifically, when $d_i = d_j = d$ (Figure 7B, left), these decision surfaces are planes of constant $\Psi_i - \Psi_j$ (=$\mathcal{C}/d$). 

Note that in the m-ADC model the decision is based on the relative magnitudes of $\Psi_i - c_i \gtreqless \Psi_j - c_j$ (\ref{madc:DR}); this can be rearranged as $\Psi_i - \Psi_j \gtreqless c_i - c_j$ or $\Psi_i - \Psi_j \gtreqless b_{ij}$ where $b_{ij} = c_i - c_j$ is the bias for location $j$ relative to locations $i$, quantified as the difference of the criteria between the two locations. Thus, when $d_i = d_j$ the decision surfaces in the m-ADC model (constant $\Psi_i - \Psi_j$) belong to the family of optimal decision surfaces for discriminating between a stimulus at location $i$ versus at location $j$. 

Thus, the optimal decision surfaces for additive Gaussian signal and noise distributions are hyperplanes in m-dimensional perceptual space, a key feature of the m-ADC model (Figure 2C). The proof for why these surfaces should intersect at a point is beyond the scope of this article (but see \cite{middleton}). However, it is clear that if these don't intersect at point, then the perceptual space could contain domains where the optimal decision is not uniquely specified.

We summarize below the generalized log-likelihood ratios, as well as the equations for each optimal decision boundary: 
\begin{empheq}[box=\myyelbox]{align} 
\Lambda_{j0}(\mathbf{W}) & =  \log \frac{p_j}{p_0} + \Psi_j d_j - d_j^2/2 & \Psi_j &= c_j \\
\Lambda_{ij}(\mathbf{W}) & =  \log \frac{p_i}{p_j} + \Psi_i d_i - \frac{d_i^2}{2} - \Psi_j d_j + \frac{d_j^2}{2} & \Psi_i - \Psi_j &= c_i - c_j 
\end{empheq}

These results demonstrate that our choice of decision surfaces in the m-ADC model is optimal (for maximizing average utility or minimizing risk), when the sensitivities ($d_i$) across locations are identical. Figure 2C illustrates these optimal decision surfaces for the 2-ADC case (thick black lines).

\subsection*{Psychometric functions in the m-ADC model}
\addcontentsline{toc}{subsection}{Psychometric functions in the m-ADC model}

The m-ADC model developed so far provides a means to estimate perceptual sensitivity while controlling for choice bias when the stimulus was either presented at a fixed strength at one of the $m$ locations or absent. However, perceptual sensitivity ($d$) is, in fact, a function of stimulus strength: stronger, more salient stimuli are more reliably detected because the signal distribution for such stimuli is further removed (translated) from the noise distribution (higher $d$), resulting in decreased overlap between the signal and noise distributions (Figure 2A). The \textit{psychometric function} describes the variation of perceptual sensitivity $d$ with stimulus strength. 

It is often of interest to understand how an experimental manipulation, such as cueing a particular location for attention, affects the underlying psychometric function: does the manipulation scale, shift or change the slope of the psychometric function? Such an analysis is fundamental to evaluating competing hypotheses (normalization, response gain, contrast gain) regarding the nature of attention's effects on perception \cite{reynoldsheeger2009, hermannheeger2010, leemaunsell2009}. 

In this section, we extend the m-ADC model to estimate the psychometric function of detection sensitivity. We demonstrate that the various effects on the psychometric function (scaling, shift, slope change) can be directly inferred from the response probabilities with a one-shot estimation procedure. We demonstrate that models that do not account for bias (differences in $c$) across locations, produce spurious differences in psychometric function estimates. Finally, we demonstrate that ignoring catch trials in the analysis, and treating the behavioral data as if they were acquired in a forced choice design, also results in systematically biased psychometric function estimates. 

\subsubsection*{Relating response probabilities to the psychometric function}
\addcontentsline{toc}{subsubsection}{Relating response probabilities to the psychometric function}

In order to account for the variation of perceptual sensitivity ($d$) with stimulus strength, we introduce the following modification to the structural model:
\begin{equation} \label{eq:cmod}
\Psi_i  =  d_i(\xi_i) + \varepsilon_i 
\end{equation}

\noindent 
where $\xi_i, i \in \{1, \ldots, M\}$ represents some metric of stimulus strength at location $i$, and the psychometric function $d_i(\xi_i)$ describes the functional dependence of sensitivity at location $i$ on the stimulus strength at that location. For ease of description, we choose $\xi_i$ to represent the contrast of the stimulus. In this exemplar case our theory relates response probabilities to the well-known psychometric function of stimulus contrast. 

In addition, we assume that the subject employs a fixed criterion, $c_i$, at each location that is independent of (does not vary with) stimulus strength. Such an assumption is plausible for task designs in which stimulus strength is varied pseudorandomly across trials, so that the subject is unaware of the strength of an upcoming stimulus and, hence, cannot adjust her/his criterion systematically with stimulus strength. Thus, the decision rule remains the same as the m-ADC decision rule (\ref{madc:DR}).

Based on equations (\ref{eq:maufc}) and (\ref{eq:cmod}), response probabilities for each stimulus contrast $\xi$ may be derived as:
\begin{empheq}[box=\myyelbox]{align} 
p(Y = i | \ \boldsymbol{\xi}) &= \int_{c_i - d_i(\xi_i)}^{\infty} \prod_{k, k \neq i}  F_k(e_i + d_i(\xi_i) - d_k(\xi_k) - c_i + c_k) \ f_i(e_i) \ de_i \label{eq:psymaufc} 
\end{empheq}

\noindent
where $\boldsymbol{\xi}$ represents the contrast of the stimulus presented at each of the m-locations $(\xi_1, \xi_2, \ldots, \xi_m)$. In the m-ADC task design, the stimulus may be presented at no more than one location on a given trial so that all but one of the $\xi_i$ are zero.

In order to model the psychometric function, we look for a parametric function such that the parameters of this function independently control the scale, shift or slope of the function. A variety of parametric functions, such as the hyperbolic-ratio function, logistic function or Weibull function have been used in previous studies. For illustration, we choose the commonly used hyperbolic-ratio (or Naka-Rushton) function; the results that follow hold regardless of the specific choice of function.

The hyperbolic-ratio function, defined as: 
\begin{equation}
d(\xi) = d_{max} \left(\frac{\xi^n}{\xi^n + {\xi_{50}}^n}\right) \label{eq:NRfunc}
\end{equation}
provides an analytical relationship between sensitivity $d$, and stimulus contrast, $\xi$. The parameters of this function, $d_{max}$, $\xi_{50}$ and $n$ (which we call \textit{psychometric parameters}) correspond to the asymptotic value, contrast at 50\% of asymptotic value, and slope of the psychometric function, respectively. Altering each parameter in turn scales ($d_{max}$), shifts ($\xi_{50}$) or changes the slope ($n$) of the psychometric function. Note that, by definition, altering the criteria should have no impact on the psychometric function. 

\subsubsection*{Estimating psychometric parameters from response probabilities}
\addcontentsline{toc}{subsubsection}{Estimating psychometric parameters from response probabilities}

As before, we demonstrate parameter recovery with the 2-ADC model. The demonstration is readily extended to the m-ADC case.

For the 2-ADC model, the response probabilities may be written, by simplifying (\ref{eq:psymaufc}) as:
\begin{align} \label{eq:psy2aufc}
p(Y = i |\ \xi_i, \xi_j)& = \int_{c_i - d_i(\xi_i)}^{\infty}  F_j(e_i + d_i(\xi_i) - d_j(\xi_j) - (c_i - c_j)) \ f_i(e_i) de_i \\ \nonumber
&\footnotesize{i,j \in \{1, 2\}}
\end{align}

The psychometric function of contrast at each location ($d_i(\xi_i)$) is defined by a set of parameters specific to that location $(d_{max}, \xi_{50}, n)_i$ (Table S3). Figure S1 depicts the effect of varying each psychometric parameter and choice criterion on the response probabilities, for stimuli of various contrasts at location 1, $p(Y=i|\xi_1)$ (parameter values in Table S3A). 

For values of the parameters that do not saturate the response probabilities, the effect of varying the psychometric parameters $d_{max}$, $\xi_{50}$ and $n$ on the response probability curves (Figure S1A-C) is similar to the effect of the respective parameter on the psychometric function, $d(\xi)$, viz. scaling, shift and slope change (Figure S1A-C, insets). On the other hand, altering each response criterion ($c_i$) alters the response probability curves in complex ways: the effects include apparent scaling, shift and/or slope changes (Figure S1 D-E). However, the response probability at a location increases across all $\xi$ with decreasing criterion at that location, and with increasing criterion at the opposite location, consistent with the monotonic trends noted before (Figure 5A).

Psychometric parameters $(d_{max}, \xi_{50}, n)_i$ and biases $c_i$ at each location could be reliably recovered from simulated response probabilities with a one-shot ML estimation procedure. As before, response probabilities were computed by simulating equation (\ref{eq:psy2aufc}) based on a prespecified set of bias and sensitivity parameters (Table S3A) at six equally-spaced values of contrast ($\xi_k \in [0,100]$), with 50\% catch trials and 25\% stimulus trials at each of the two locations; this process was repeated for 100 simulated runs (1000 trials per contrast value for each simulation). As before, we denote these by $\mathcal{P}^r_s(\xi_k)$ (Figure 7-B circles, errorbars denote standard deviations across simulated runs), corresponding to the probability of response at location $r$ when a stimulus is presented at location $s$ with contrast $\xi_k$ ($k$ = 1-6). These simulated response proportions were provided as input to a maximum likelihood estimation algorithm in lieu of experimental data. 

Psychometric parameters and biases were reliably recovered with the ML estimation procedure (Table S3B). Response probability curves, as well the psychometric functions, computed from the recovered parameters fit the data with virtually no error (Figure 6A-B, solid curves, Figure 6C solid black curve).

\subsubsection*{Effects of ignoring choice bias or catch trial/NoGo performance}
\addcontentsline{toc}{subsubsection}{Effects of ignoring choice bias or catch trial/NoGo performance}

\begin{figure}[!t]
\begin{center}
\includegraphics[width=6.35in]{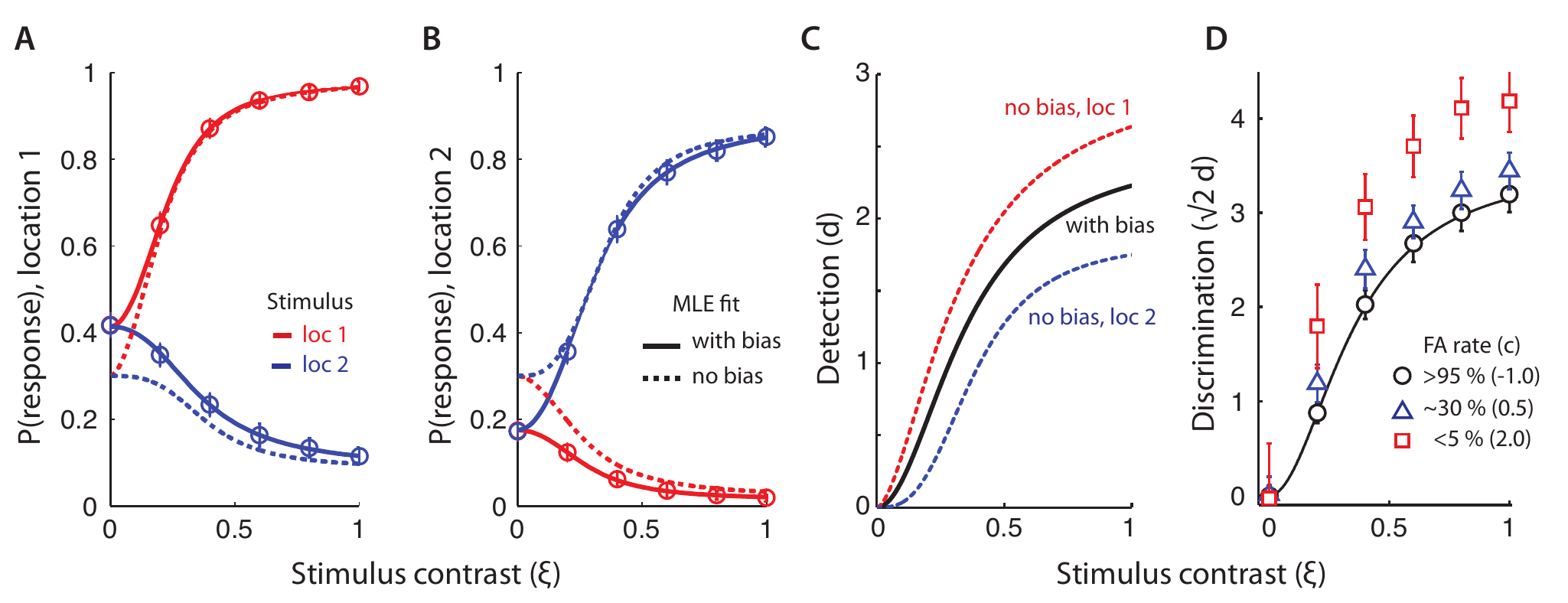}
\end{center}
\caption{
{\bf Psychometric functions in the 2-ADC model.}  
({\bf A}) Probability of response at location 1 as a function of the contrast of a stimulus at location 1 (red circles) or location 2 (blue circles). Error-bars: Standard deviation across simulated runs (N=100). Solid curves: Maximum likelihood estimate (MLE) fits to response probabilities with a model that incorporated bias. Dashed curves: MLE fits with a model that did not incorporate bias. 
({\bf B}) Same as in (A), but response probability at location 2.
({\bf C}) Maximum likelihood estimates of the psychometric function of detection sensitivity, $d(\xi)$, as a function of stimulus contrast. Black curve: Psychometric function estimated with a model that incorporated bias. Red and blue dashed curves: Psychometric functions at locations 1 and 2, respectively, estimated with a model that did not incorporate bias. 
({\bf D}) Discrimination accuracy as a function of stimulus contrast estimated by excluding catch trial/NoGo performance. Black curve: 2-ADC psychometric function of discrimination accuracy ($\sqrt{2} d(\xi)$) estimated incorporating catch trials and NoGo responses. Symbols: 2-AFC psychometric functions of discrimination accuracy estimated excluding catch trials and NoGo responses for three different cases: i) low criteria and high false-alarm rates ($c=-1.0$, FA > 95\%, black circles), ii) intermediate values of criteria and false-alarm rates ($c=0.5$, FA $\sim$30\%, blue triangles), and iii) high criteria and low false-alarm rates ($c=2.0$, FA < 5\%, red squares).   
}
\label{fig_psyc}
\end{figure}

In the previous simulations, we employed identical psychometric parameters at the two locations resulting in identical psychometric function estimates. However, the choice criteria at the two locations were specified to be different, with a higher criterion at location 2 ($c_1 = 0.1, c_2 = 0.7$), implying a choice bias toward location 1. What are the consequences of ignoring this bias (differences in $c_i$-s across locations) for estimating the psychometric function? 

We estimated values of the psychometric function based on a model that ignores bias by assuming uniform $c_i = c, i \in \{1, 2\}$ across the two locations. The resultant psychometric parameter estimates and psychometric function are shown in Table S3B and Figure 6C (dashed curves) respectively. Even though the reconstructed response probabilities (Figure 6A-B, dashed curves) closely matched the original data (Figure 6A-B, circles), detection sensitivities $d_i$ were systematically overestimated at location 1, the location of greater bias (Figure 6C, dashed red curve), and systematically understimated at the other location (Figure 6C, dashed blue curve). 

In the m-ADC model, the perceptual sensitivity $d$, for detecting a stimulus at a location, is also related to the discrimination accuracy (the accuracy with discriminating between two locations of stimulus presentation, as in a 2-AFC task) by a scale factor of $\sqrt{2}$ (Figure 2C, distance between centers of red and blue distributions, assuming $d_1 = d_2 = d$; see also Appendix in \cite{decarlo}).  Thus, the 2-ADC psychometric function of discrimination accuracy is also related to the psychometric function of detection accuracy $d(\xi)$ by a scale factor of $\sqrt{2}$. 

We estimated discrimination accuracy across the two locations by treating the data as if they were acquired with a two-alternative forced choice (2-AFC) task. For this analysis we ignored catch trials and NoGo responses, considering observations from only the first two rows and columns of the contingency table (Table S1B). In addition, for ease of demonstration, this analysis was performed with a model incorporating equal criteria across the two locations $c_i = c, i \in \{1, 2\}$, although similar results hold if the criteria are different. 

For low values of the criteria ($c =-1.0$), corresponding to high false alarm rates during catch trials (>95\%), 2-AFC estimates of discrimination accuracy (Figure 6D, black circles) matched the true (2-ADC) psychometric function of discrimination accuracy (Figure 6D, solid black curve, obtained by multiplying black data in Figure 6C by $\sqrt{2}$). For intermediate values of $c$ ($c=0.5$, false alarm rates $\sim$30\%), 2-AFC estimates of discrimination accuracy were slightly greater (Figure 6D, blue triangles) than the true value (black curve). For high values of $c$ ($c=2.0$) and very low false alarm rates during catch trials (<5\%), 2-AFC estimates of discrimination accuracy substantially exceeded (Figure 6D, red squares) the true value (black curve). Note that in each case, the 2-AFC accuracy estimates were derived from all relevant observations in the data (excluding catch trials and NoGo responses) and, hence, fit these data perfectly.  

Thus, despite appearing to fit the observed data accurately, models that do not account for choice bias or those that ignore catch trial/NoGo performance produce highly erroneous, systematically inaccurate estimates of perceptual sensitivity.


\section*{Discussion}
\addcontentsline{toc}{section}{Discussion}

With the growing use of multialternative tasks for investigating the neural basis of perceptual and cognitive phenomena, the need for developing new analytical models and theoretical frameworks for such tasks is being increasingly recognized \cite{churchlandditterich, niwaditterich}. In this study, we have developed a theoretical model that decouples the effects of choice bias from those of perceptual sensitivity in multialternative detection tasks. We demonstrate a unique and optimal mapping between model parameters and response probabilities, and provided numerical methods for estimating model parameters rapidly and reliably. Thus, our model provides a powerful tool for analyzing behavioral performance in widely used multialternative tasks of perception, attention and decision-making. 

Simulating the m-ADC model revealed several important caveats when analyzing and interpreting multialternative behavioral data. When analyzed with a model that did not incorporate choice bias, despite close fits to simulated behavioral data (Figure 6A-B dashed lines vs. circles), maximum-likelihood estimation produced large and systematic differences in psychometric functions of sensitivity across locations whereas the actual underlying psychometric functions at the two locations were, in fact, identical (Figure 6C, dashed red and blue vs. solid black curve). These spurious differences arose because a greater proportion of responses arising from an increased choice bias to a location were erroneously attributed to increased perceptual sensitivity at that location. Furthermore, estimating the psychometric function after excluding catch trials and NoGo responses resulted in systematically inflated estimates of discrimination accuracy, particularly when false-alarm rates were low. This result has important implications for multialternative behavioral designs that report low false alarm rates during catch trials as evidence against a guessing strategy, and ignore data from these catch trials in the subsequent analysis of behavioral responses. 

In any behavioral model, demonstrating the uniqueness of underlying model parameters is necessary to meaningfully interpret the behavioral significance of absolute (or relative) parameter values \cite{bruntonbrody}.  The m-ADC model is among the most parsimonious class of analytical models for multialternative detection as a result of several key assumptions (discussed next); this parsimony permitted us to analytically demonstrate the uniqueness of the sensitivity and bias parameters for multialternative tasks with any number of alternatives (models of arbitrarily high dimensions). Such an analytical demonstration is considerably more challenging, and often never accomplished, for more complex models.

\subsection*{Assumptions and extensions of the m-ADC model}
\addcontentsline{toc}{subsection}{Assumptions and extensions of the m-ADC model}

The m-ADC latent variable formulation makes several assumptions: (i) underlying perceptual dimensions (decision variable axes) are orthogonal; (ii) distributions of the decision variables are independent (covariance matrix of $\mathbf{\Psi}$ is diagonal); (iii) variance of the decision variable distribution does not change with stimulus contingency; and (iv) decision boundaries are linear (or planar). In the following, we discuss which of these assumptions are reasonably justified, and which can be addressed by extending the model.

The assumption of independent decision variable distributions that are represented along orthogonal dimensions (independent channels) has been tested in the two-dimensional case, and found valid for stimulus attributes that are widely different perceptually (for instance, stimuli that are widely separated in space or frequency) \cite{tanner}. However, it is possible that the $\Psi_i$-s are not independent, and perceptual sensitivities do not vary along orthogonal dimensions: signal covariation may arise from facilitative or competitive interactions that operate across locations. In addition,  decision variable distributions at different locations could be correlated or, equivalently, decision variable axes could be separated by angles different from $90^{\circ}$. In this case the covariance matrix of $\mathbf{\Psi}$ is no longer diagonal; the model could be extended to incorporate this scenario. 
 
Equal variance for the decision variable distributions under signal and noise is a basic assumption of conventional signal detection models; such an assumption permits defining a one-to-one relationship between the likelihood ratio and the decision variable axis. This assumption can be tested in the m-ADC model. As the m-ADC model has a surplus of independent observations relative to parameters ($m^2-m)$, unequal variances of the signal distribution at each location may be incorporated into the model by introducing the signal-to-noise variance ratio at each location as independent parameters, although specifying an optimal decision rule is more complex with such models. 

We have demonstrated that, for additive Gaussian signal and noise distributions, planar hypersurfaces (hyperplanes), as defined by the choice criteria in the model, constitute a family of optimal decision surfaces. A subset of decision surfaces in the current model (of the form $\Psi_i - \Psi_j = c_i - c_j$) are optimal only if the values of sensitivity ($d$) are identical across locations ($d_i = d \ \ \forall i$, Figure 7B, left). In certain experiments, such as when a particular spatial location is cued for attention, it is possible that $d$-s at the two locations could be significantly different. The model may then be extended with a modified decision rule to capture optimal decision making in this more general scenario of unequal $d$-s (Figure 7B, right). 

Although not an assumption of the model, our task specification requires that no more than one stimulus be presented on a given trial. A particular advantage of this task specification is that potential second and higher interaction terms (of the form $X_i X_j, X_i X_j X_k, \ldots$) in the structural model vanish automatically (as at least one $X_i = 0$). Tasks that violate this requirement and incorporate compound stimulus contingencies (e.g. stimuli presented at more than one location, or more than one stimulus feature presented, on a given trial) fall under the purview of the General Recognition Theory (GRT, \cite{ashbybk}) framework (discussed next).

\subsection*{Relationship to previous signal detection models}
\addcontentsline{toc}{subsection}{Relationship to previous signal detection models}

A variety of two (and higher) dimensional signal detection (SDT) models have been proposed in literature \cite{mcsdtbk}: these are generally referred to as multichannel SDT models. However, the analysis of multialternative detection tasks poses unique challenges, because such tasks fall under the category of partial identification designs (see below); accounting for bias in SDT models for multialternative detection tasks remains an important, open problem \cite{mcsdt10} (pp. 258, Figure 10.4). 

The m-ADC model accounts for bias by incorporating an asymmetric decision rule (unequal criteria) building upon a recently developed latent variable formalism for dealing with bias in multialternative forced-choice (m-AFC) tasks \cite{decarlo}. How are the m-ADC and m-AFC models related?

The SDT model for a 2-AFC (binary choice) task involves a decision based on distinguishing between two decision variable distributions. In a forced choice (2-AFC) task, these are the two signal distributions, one for each hypothesis (e.g. stimulus at location 1 vs. 2, Figure 7A), whereas in a simple detection (Yes/No) task, these are the signal (stimulus present) and noise (stimulus absent) distributions. On the other hand, a two-alternative detection (2-ADC) task involves discriminating among three distributions: two signal distributions, one representing each stimulus hypothesis, and a noise distribution (catch trials, Figure 7B). For the 2-AFC task, the criterion represents a decision surface to distinguish between the two signal distributions (Figure 7A), whereas in the 2-ADC task, the two criteria represent decision boundaries between the noise distribution and each of the signal distributions (Figure 7B).

Based on this description, it is apparent that the m-ADC model reduces to the m-AFC model if catch trials and NoGo responses are excluded (see also formal derivation in Appendix B, Supporting Information). Thus, behavioral data that can be fit well with the m-AFC model can also be fit with the m-ADC model. In general, the m-ADC design (with catch trials) also has the advantage of having more independent observations than parameters ($m^2-m$) relative to the m-AFC design ($m^2-3m+1$). The additional independent observations permit more complex models, with more parameters to be specified, and also permits rigorous internal tests of validity relative to the m-AFC model. Moreover, unlike the m-AFC task design the m-ADC task design permits measuring true detection sensitivities for each of the m-alternatives and overcomes potential confounds associated with a guessing strategy.

\begin{figure}[!t]
\begin{center}
\includegraphics[width=5.3in]{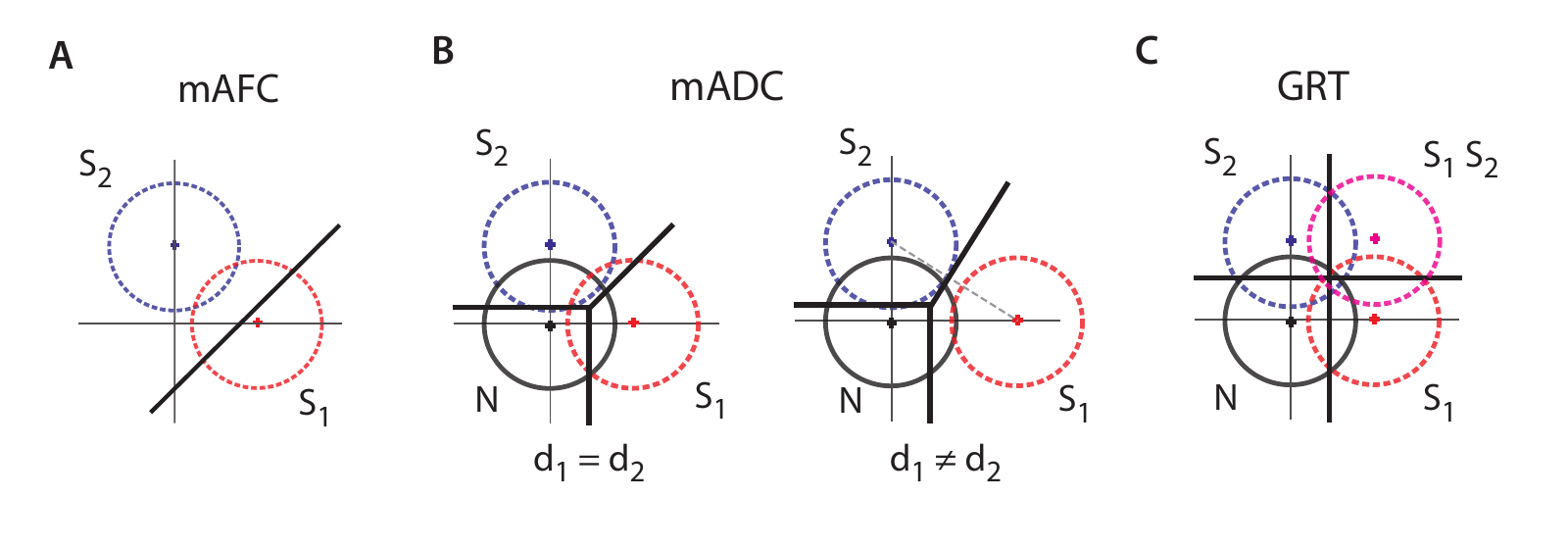}
\end{center}
\caption{
{\bf Multidimensional signal detection models for analyzing multialternative task designs.}  
({\bf A}) Model schematic for a multialternative forced choice (here, 2-AFC) design \cite{decarlo}. (All panels) Circles: decision variable distributions for each stimulus contingency. Red: contingency $S_1$ (e.g., stimulus at location 1); blue: contingency $S_2$ (e.g, stimulus at location 2). Thick black line: Optimal decision surface. 
({\bf B}) Model schematic for a multialternative detection (here, 2-ADC) design. In addition to stimulus contingencies $S_1$ and $S_2$ a third no-stimulus contingency, $N$ (black circle) occurs during ``catch'' trials. Left and right panels: Decision variable distributions and optimal decision surfaces for equal (left) or unequal (right) sensitivities across contingencies. 
({\bf C}) Model schematic for a complete identification design, based on general recognition theory (GRT) \cite{ashbybk}. In addition to contingencies $S_1$, $S_2$ and $N$ a fourth compound stimulus contingency, $S_1 S_2$ (dashed gray circle) occurs in trials where both stimuli (or stimulus features) are presented. In the spatial detection task (such as in Figure 1), this would correspond to the presentation of a stimulus (or change) at more than one location on a given trial.
}
\label{fig_model}
\end{figure}

Our m-ADC model follows a rich literature on multidimensional signal detection models within the framework of general recognition theory (GRT) \cite{ashbybk}. In the psychoacoustic and vision literature, GRT models have been widely applied in tasks involving the detection of multiple signals in noise \cite{ashbybk, ashbytownsend}. These models are relevant for tasks that implement a feature complete identification design (\cite{mcsdt10} pp.260), such as those shown in Figure 7C. This task involves discriminating four potential stimulus contingencies (Figure 7C): noise alone ($N$), each stimulus alone ($S_1$ or $S_2$), or the compound stimulus ($S_1 S_2$). Such a four-way ($2 \times 2$) discrimination simplifies the optimal decision rule (orthogonal pairs of lines) for Gaussian signals and noise (Figure 7C, thick black lines) \cite{ashbytownsend}. In the m-ADC tasks the stimulus (or change) occurs at no more than one location on a given trial. Thus, the last contingency (compound, $S_1 S_2$) of a GRT design is never presented. Thus, the GRT model and decision rule do not apply to the m-ADC task.

A variety of models for dealing with bias in multialternative tasks have been formulated within the framework of Luce's choice theory \cite{luce1963}; standard methods in textbooks of behavioral analysis account for bias with a choice theory model (\cite{mcsdt10} pp.250). In contrast, few attempts have been made to deal with bias in multialternative tasks from a signal detection framework \cite{decarlo}. Early attempts at two-dimensional ``detection and recognition'' or (``detection and identification'') models began with \cite{tanner, swetsbirdsall}; although conceptually similar to the m-ADC model, these models were geometrically formulated. Later studies attempted to develop a mathematical formalism for these models by treating the decision variable as a random vector akin to the multivariate decision variable in the m-ADC model \cite{thomasolzak}; these models were formulated for double-judgment (detection and identification) tasks. The importance of accounting for bias to avoid spurious conclusions in multidimensional models for such double judgment tasks has been clearly highlighted \cite{klein}. However, these earlier formulations were based on complete identification designs, those that incorporate the compound stimulus contingency ($S_1 S_2$, Figure 7C) \cite{thomasolzak, olzakthomas}. 

On the other hand, psychophysical tasks of detection and attention, such as those presented in this study (Figure 1) and elsewhere \cite{cavanaughwurtz,cohenmaunsell}, are all partial (or incomplete) identification designs in which no more than one stimulus or change is presented on a given trial (Figure 7B). Hence, although the mathematical formalism in previous SDT models resembles the m-ADC model, the decision rule is significantly different (Figure 7B vs. 7C). The specification of the m-ADC decision rule is, arguably, more complex and, consequently, the analytical formulation of bias in multialternative detection tasks is a fundamentally novel aspect of the m-ADC model. 

\subsection*{Conclusion}
An animal's behavior results from an amalgamation of various factors: perceptual, motivational, decisional, and the like. Inferring the complex relationship between an animal's perception and its behavior is currently best accomplished by recourse to theoretical frameworks \cite{carandinichurchland}. Frameworks, such as signal detection theory, capture the effect of perceptual (sensitivity) vs. decisional (bias) factors at the level of the organism as a behavioral unit. The m-ADC model developed in this study provides a rigorous and testable framework of how sensitivity and bias affect the animal's behavior in multialternative detection tasks. Future work will involve extending this model to incorporate the influences of executive and cognitive processes (such as attention) to sensitivity and bias as well as validating and refining the model with behavioral data.


\clearpage
\section*{Methods}
\addcontentsline{toc}{section}{Methods}

\subsection*{Linking sensitivities and criteria to 2-ADC response probabilities}
\addcontentsline{toc}{subsection}{Linking sensitivities and criteria to 2-ADC response probabilities}

In the 2-ADC model, the probability of response at each location ($Y = i$) for each possible stimulus condition ($X_i$) can be derived from the structural model (\ref{eq:strmod}) and decision rule (\ref{eq:decrule}). We illustrate the case for $p(Y = 1|X)$. The other cases may be similarly derived.

The probability of response at location 1 is the combined probability that the decision variable at location 1 exceeds the choice criterion at that location, and that its magnitude (over its choice criterion) is the larger of the two locations.
\begin{equation}
p(Y = 1 | X_1, X_2) =  p(\Psi_1 > c_1 \cap \Psi_1 - c_1 > \Psi_2 - c_2) \\
\end{equation}

\noindent 
which, upon substitution of the structural model gives: 
\begin{equation}
\begin{split}
p(Y = 1 | X_1, X_2)&= p(\varepsilon_1 > c_1 -d_1 X_1 \ \cap \\
	&\varepsilon_1 + d_1 X_1 - c_1 > \varepsilon_2 + d_2 X_2 - c_2) \\
\end{split}
\end{equation}

\noindent
We condition the above probability on a given value of $\varepsilon_1 = e_1$. 
\begin{align}
 &p(Y = 1 | X_1, X_2, \varepsilon_1 = e_1) \\
 &= p(e_1 > c_1 - d_1 X_1 \ \cap e_1 + d_1 X_1 - c_1 > \varepsilon_2 + d_2 X_2 - c_2) \nonumber \\
 &= \mathcal{H}(e_1 - c_1 + d_1 X_1) \ p(\varepsilon_2 < e_1 + d_1 X_1 - d_2 X_2 - c_1 + c_2) \nonumber \\
 &= \mathcal{H}(e_1 - c_1 + d_1 X_1) \ F_2(e_1 + d_1 X_1 - d_2 X_2 - (c_1 - c_2)) 
\end{align}

where $\mathcal{H}(x)$ is the Heaviside function and $F_2$ represents the cumulative distribution function (CDF) of the noise distribution at location 2, $\varepsilon_2$.

\par
The conditional probability for a response at location 1 is found by integrating over the distribution of $e_1$.
\begin{equation}
\begin{split}
p(Y = 1 | X_1, X_2) &= \int_{-\infty}^{\infty} \mathcal{H}(e_1 - c_1 + d_1 X_1) \\
& F_2(e_1 + d_1 X_1 - d_2 X_2 - (c_1 - c_2)) \ f_1(e_1) \ de_1
\end{split}
\end{equation}

\noindent
where $f_1$ represents the probability density function of the noise distribution at location 1, $\varepsilon_1$.

\noindent

The Heaviside function may be dropped from the integrand by defining the lower bound of the integral at $c_1 - d_1 X_1$. In other words, 
\begin{equation} \label{eq:sixindep1} 
\begin{split}
p(Y = 1 | X_1, X_2)&= \int_{c_1 - d_1 X_1}^{\infty}  F_2(e_1 + d_1 X_1 - d_2 X_2 \\ 
& \qquad \qquad \quad {} - (c_1 - c_2)) f_1(e_1) \ de_1
\end{split}
\end{equation}

\noindent
Similarly, the conditional probability of a response at location 2 is given by: 
\begin{equation} \label{eq:sixindep2} 
\begin{split}
p(Y = 2 | X_1, X_2)&= \int_{c_2 - d_2 X_2}^{\infty}  F_1(e_2 + d_2 X_2 - d_1 X_1 \\
& \qquad \qquad \quad {} - (c_2 - c_1)) \ f_2(e_2) \ de_2
\end{split}
\end{equation}

\noindent
These equations represent six independent observations corresponding to the Go response ($Y = 1$ or $Y = 2$). Three other response probabilities, corresponding to the NoGo response ($Y = 0$) are not free to vary as the Go and NoGo responses are mutually exclusive and exhaustive. These probabilities are readily calculated as: $p(Y = 0| X_i) = 1 - p(Y = 1| X_i) - p(Y = 2| X_i)$, and can be shown to be equal to $\prod_{i=1}^2 F_i(c_i - d_i X_i)$. Each of these nine probabilities corresponds to one of the nine contingencies in the 2-ADC contingency table (Table S1B).

The set of equations (\ref{eq:sixindep1}) and (\ref{eq:sixindep2}) represents an overdeterminate system, as there are six independent observations, but only four parameters \{$d_i, c_i$\} (i = 1, 2). Thus, there is redundancy in the response probabilities (only four are free to vary, given the model parameters). These remaining probabilities can then be used as an internal test of the validity of the model. 

As equations (\ref{eq:sixindep1}) and (\ref{eq:sixindep2}) have similar forms, we will combine these into a single system of equations. These, together with the equations describing the NoGo response probabilities, constitute the 2-ADC model system (reproduced in the Results as equation system (\ref{eq:sixindep3})).
\begin{equation} 
\begin{split}
p(Y = i | X_i, X_j)&= \int_{c_i - d_i X_i}^{\infty}  F_j(e_i + d_i X_i - d_j X_j \\
& \qquad \qquad \qquad {} - (c_i - c_j)) \ f_i(e_i) \ de_i \\ 
\end{split}
\end{equation}
\vspace{-20pt}
\begin{align*}  
p(Y = 0 | X_i, X_j) &=  F_i(c_i - d_i X_i) \ F_j(c_j - d_j X_j) & \\
& \footnotesize{i,j \in \{1,2\}, i \neq j} 
\end{align*}

In simulations, we evaluated these probabilities with numerical integration. As the normal distribution has infinite support, the integrands on the right hand sides of these equations should be integrated to an upper limit at plus infinity, a numerically intractable bound. We used Gauss-Kronrod quadrature (as implemented in the \textit{quadgk} function in Matlab) in order to evaluate these integrals.

\subsection*{Linking sensitivities and criteria to m-ADC response probabilities}
\addcontentsline{toc}{subsection}{Linking sensitivities and criteria to m-ADC response probabilities}

Based on the structural model (\ref{madc:SM}) and decision rule (\ref{madc:DR}) in the m-ADC model we derive the probabilities of response at location $i$ ($Y = i$) given a stimulus at location $j$ ($X_j = 1$).
\begin{equation}
\begin{split}
p(Y = i | \mathbf{X})& =  p((\Psi_i > c_i) \ \cap \ (\Psi_i - c_i > \Psi_1 - c_1) \ \cap \ \\
&(\Psi_i - c_i > \Psi_2 - c_2) \ldots \ \cap \ (\Psi_i - c_i > \Psi_m - c_m) ) 
\end{split}
\end{equation}

\vspace{-5pt}
\noindent
where we have used the notation $\mathbf{X}$, which represents a vector $(X_1, X_2, \ldots, X_m)$. For a stimulus presented at the j-th location ($X_j = 1$) the j-th element of the vector is 1 and all other elements 0; when no stimulus is presented (catch trial) all elements are zero ($X_i = 0 \ \ \forall \  i$). Thus, $\lVert\mathbf{X}\rVert_{1} = 1$ (stimulus trial) or $0$ (catch trial); this follows from the task specification, that no more than one stimulus is presented on a given trial. 

Upon substitution of the structural model, this gives: 
\begin{align}
p(Y = i | \mathbf{X})& = p((\varepsilon_i > c_i - d_i X_i) \ \cap_k \ \\ \nonumber
& (\varepsilon_i + d_i X_i - c_i > \varepsilon_k + d_k X_k - c_k) & i \neq k 
\end{align}

Similar to the 2-ADC case, we condition the above probability on a given value of $\varepsilon_i = e_i$. 
\begin{equation}
\begin{split}
p(Y = i | \mathbf{X}, e_i) &= p((e_i > c_i - d_i X_i) \ \cap_{k,  k \neq i} \\ 
& \qquad (e_i + d_i X_i - c_i > \varepsilon_k + d_k X_k - c_k)  \\
&= \mathcal{H}(e_i - c_i + d_i X_i) \ \cap_{k,  k \neq i} \\ 
& \qquad p(\varepsilon_k < e_i + d_i X_i - d_k X_k - c_i + c_k) \\
&= \mathcal{H}(e_i - c_i + d_i X_i) \\
& \qquad \prod_{k, k \neq i}  F_k(e_i + d_i X_i - d_k X_k - c_i + c_k)
\end{split}
\end{equation}

\vspace{-5pt}
\noindent
where $\mathcal{H}(x)$ is the Heaviside function and $F_k$ represents the cumulative distribution function of the decision variable distribution at location $k, \Psi_k$. In deriving this expression, we have used the fact that the $\Psi_k$ distributions are mutually independent, such that their joint probability density factors into a product of the individual densities.

The probability of a response at location $i$ is then found by integrating over the probability density of $e_i$.
\begin{equation}
\begin{split}
p(Y = i | \mathbf{X}) &=  \int_{-\infty}^{\infty} \mathcal{H}(e_i - c_i + d_i X_i) \ \\ 
& \prod_{k, k \neq i}  F_k(e_i + d_i X_i - d_k X_k - c_i + c_k) \ f_i(e_i) \ de_i \\
&= \int_{c_i - d_i X_i}^{\infty} \prod_{k, k \neq i}  F_k(e_i + d_i X_i \\
& \qquad \qquad \quad {} - d_k X_k - c_i + c_k) \ f_i(e_i) \ de_i 
\end{split}
\end{equation}

This constitutes the m-ADC model system of equations relating response probabilities for each stimulus contingency to the sensitivity and criterion at each location (reproduced in the results as equation system (\ref{eq:maufc}).

\subsection*{Maximum likelihood and Markov-Chain Monte Carlo estimation of sensitivities and criteria in the m-ADC model}
\addcontentsline{toc}{subsection}{Maximum likelihood and Markov-Chain Monte Carlo estimation of sensitivities and criteria in the m-ADC model}

The maximum likelihood (line-search) or ML-LS algorithm was implemented by minimizing the negative of the log-likelihood function with an unconstrained minimization algorithm (\textit{fminunc}, in Matlab's Optimization Toolbox). The optimization algorithm also returns a numerical approximation to the Hessian matrix. Standard errors based on ML-LS estimation were derived as the square root of the diagonal elements of the inverse of this Hessian matrix. Our algorithm for ML estimation differs from the previously published algorithm for the related m-AFC task, where each response variable was assumed to follow an independent Bernoulli distribution \cite{decarlo}. The responses in our task were assumed to follow a multinomial distribution. For both m-ADC and m-AFC tasks, conditional response probabilities are not independent: the total number of responses for each stimulus contingency sums to the number of trials for that contingency; thus, the conditional response probabilities for each stimulus contingency should sum to one.

The Markov Chain Monte Carlo algorithm (MCMC, Metropolis sampling) was custom implemented for estimating sensitivity and criteria from simulated response counts $\mathcal{O}^r_s$ (denoting the number of responses to location $r$ for a stimulus at location $s$). In the following, $N_s$ denotes the total number of trials for each stimulus contingency $s$ and the symbol $\mathbf{d}_i$ is used as a general notation either for sensitivity, $d_i$ when estimation was performed at a single value of stimulus strength, or for the collection of psychometric parameters $(d_{max}, n, \xi_{50})_i$ when estimation was performed on the entire psychometric function.

\begin{enumerate}[(i)] 
 \item Generate an initial guess for the parameters $(\mathbf{d}_i^r, c_i^r)$ (the superscript $r$ denotes a reference set). Designate this as the reference parameter set. Determine response probabilities from (\ref{eq:sixindep3}) based on this set. We denote these probabilities by $\varphi^r_s(\xi_k|\ \mathbf{d}_i^r, c_i^r)$. 
 \item Compute the likelihood value $\mathcal{L}^r$ assuming that responses $\mathcal{O}^r_s$ follow a multinomial distribution with parameters $N_s, \varphi^r_s$
 \item Generate a new guess for the parameters $(\mathbf{d}_i^{n}, c_i^{n})$, based on a transition probability distribution for the parameters.
 \item Determine response probabilities and the associated likelihood value $\mathcal{L}^{n}$ based on the new guess. 
 \item Compute a likelihood ratio based on the older and newer guesses: $\mathcal{L}_R = \mathcal{L}^{n}/\mathcal{L}^{r}$
 \item Accept the new guess for the parameters with a probability $a$, that depends on the magnitude of the likelihood ratio, $a=\mathrm{min}(\mathcal{L}_R, 1)$. Once accepted, the new set of parameters become the reference set, and the likelihood value based on the last set of accepted parameters is used as the reference value ($\mathcal{L}^r$). 
 \item Repeat steps (iii) - (vi) until convergence.
\end{enumerate}

We used Metropolis sampling of parameter space based on a symmetric, multivariate transition probability distribution (Gaussian, with standard deviation, $\sigma = 0.02$ in each dimension). The MCMC simulation proceeded until the algorithm converged on a specific set of parameters  ${\mathbf{d}_i, c_i}, i \in \{1,2\}$ in four-dimensional space. The algorithm was determined to have converged when the value of $\mathcal{L}$ and the $\chi^2$ error function changed by less than 2\% over at least 100 consecutive iterations. The burn-in period was generally achieved within about 500 iterations (e.g. Figure 4E-F). Posterior distributions were computed based on parameter values between iterations 1000-2000. Standard errors for the parameters and 95\% credible intervals reported (Table S2C) were based on the standard deviation and the [2.5-97.5] percentile of the posterior distributions. 

In the numerical estimation, the parameters $\{d_i, c_i\}$ were permitted to take both positive and negative values (unconstrained optimization); no constraint was placed on their sign or magnitude. However, negative values of sensitivity parameters ($d_i$) lack physical meaning. We repeated the estimation by constraining sensitivity parameters to take only positive values (with the constrained optimization function $\textit{fmincon}$ in Matlab, or with a custom-implemented MCMC Metropolis-Hastings algorithm); this analysis yielded sensitivity estimates that matched those obtained with the unconstrained optimization approaches.

\section*{Acknowledgments}
We would like to thank Lawrence DeCarlo for sharing a reprint of his study, Lynn Olzak for useful discussions, and Alireza Soltani and Peiran Gao for comments on a preliminary version of this manuscript. This research was supported by a Stanford School of Medicine Dean's Postdoctoral Fellowship (DS), Stanford MBC IGERT Fellowship and NSF Graduate Research Fellowship (NS), NIH Grant EY014924 (TM) and NIH Grant MH094938-01A1 (EK).


\bibliography{modelrefs}

\begin{thebibliography}{10}
\providecommand{\url}[1]{\texttt{#1}}
\providecommand{\urlprefix}{URL }
\expandafter\ifx\csname urlstyle\endcsname\relax
  \providecommand{\doi}[1]{doi:\discretionary{}{}{}#1}\else
  \providecommand{\doi}{doi:\discretionary{}{}{}\begingroup
  \urlstyle{rm}\Url}\fi
\providecommand{\bibAnnoteFile}[1]{%
  \IfFileExists{#1}{\begin{quotation}\noindent\textsc{Key:} #1\\
  \textsc{Annotation:}\ \input{#1}\end{quotation}}{}}
\providecommand{\bibAnnote}[2]{%
  \begin{quotation}\noindent\textsc{Key:} #1\\
  \textsc{Annotation:}\ #2\end{quotation}}
\providecommand{\eprint}[2][]{\url{#2}}

\bibitem{goldshadlen}
Gold JI, Shadlen MN (2007) The neural basis of decision making.
\newblock Annual Review of Neuroscience 30: 535-574.
\bibAnnoteFile{goldshadlen}

\bibitem{goldbennur}
Gold JI, Law CT, Connolly P, Bennur S (2008) The relative influences of priors
  and sensory evidence on an oculomotor decision variable during perceptual
  learning.
\newblock Journal of Neurophysiology 100: 2653-2668.
\bibAnnoteFile{goldbennur}

\bibitem{mcsdtbk}
Macmillan NA, Creelman DC (2005) Multidimensional Detection Theory and
  Multi-Interval Discrimination Designs, New Jersey: Lawrence Erlbaum
  Associates, Inc., chapter~10.
\newblock Detection Theory: A User's Guide. pp. 141-268.
\bibAnnoteFile{mcsdtbk}

\bibitem{klein2001}
Klein SA (2001) Measuring, estimating, and understanding the psychometric
  function: a commentary.
\newblock Perception and psychophysics 63: 1421-1455.
\bibAnnoteFile{klein2001}

\bibitem{carpenter}
Carpenter RH, Williams ML (1995) Neural computation of log likelihood in
  control of saccadic eye movements.
\newblock Nature 377: 59-62.
\bibAnnoteFile{carpenter}

\bibitem{hanksshadlen}
Hanks TD, Mazurek ME, Kiani R, Hopp E, Shadlen MN (2011) Elapsed decision time
  affects the weighting of prior probability in a perceptual decision task.
\newblock Journal of Neuroscience 31: 6339-6352.
\bibAnnoteFile{hanksshadlen}

\bibitem{mulder}
Mulder MJ, Wagenmakers EJ, Ratcliff R, Boekel W, Forstmann BU (2012) Bias in
  the brain: a diffusion model analysis of prior probability and potential
  payoff.
\newblock Journal of Neuroscience 32: 2335-2343.
\bibAnnoteFile{mulder}

\bibitem{greenswetsbk}
Green DM, Swets JA (1988) Signal Detection Theory and Psychophysics.
\newblock Los Altos, California: Peninsula Publishing.
\bibAnnoteFile{greenswetsbk}

\bibitem{cavanaughwurtz}
Cavanaugh J, Wurtz RH (2004) Subcortical modulation of attention counters
  change blindness.
\newblock Journal of Neuroscience 24: 11236-11243.
\bibAnnoteFile{cavanaughwurtz}

\bibitem{cohenmaunsell}
Cohen MR, Maunsell JH (2009) Attention improves performance primarily by
  reducing interneuronal correlations.
\newblock Nature Neuroscience 12: 1594-1600.
\bibAnnoteFile{cohenmaunsell}

\bibitem{moorefallah}
Moore T, Fallah M (2004) Microstimulation of the frontal eye field and its
  effects on covert spatial attention.
\newblock Journal of Neurophysiology 91: 152-162.
\bibAnnoteFile{moorefallah}

\bibitem{raymaunsell}
Ray S, Maunsell JH (2010) Differences in gamma frequencies across visual cortex
  restrict their possible use in computation.
\newblock Neuron 67: 885-896.
\bibAnnoteFile{raymaunsell}

\bibitem{zenonkrauzlis}
Zenon A, Krauzlis RJ (2012) Attention deficits without cortical neuronal
  deficits.
\newblock Nature 489: 434-437.
\bibAnnoteFile{zenonkrauzlis}

\bibitem{mcsdt10}
Macmillan NA, Creelman DC (2005) Multidimensional Identification, New Jersey:
  Lawrence Erlbaum Associates, Inc., chapter~10.
\newblock Detection Theory: A User's Guide. pp. 245-266.
\bibAnnoteFile{mcsdt10}

\bibitem{decarlo}
DeCarlo LT (2012) On a signal detection approach to m-alternative forced choice
  with bias, with maximum likelihood and bayesian approaches to estimation.
\newblock Journal of Mathematical Psychology 56: 196-207.
\bibAnnoteFile{decarlo}

\bibitem{luce1963}
Luce RD (1963) Detection and Recognition, New York: Wiley.
\newblock Handbook of Mathematical Psychology, Vol. 1. pp. 103-189.
\bibAnnoteFile{luce1963}

\bibitem{middleton}
Middleton D, Meter D (1955) On optimum multiple-alternative detection of
  signals in noise.
\newblock IRE Transactions on Information Theory 1: 1-9.
\bibAnnoteFile{middleton}

\bibitem{reynoldsheeger2009}
Reynolds JH, Heeger DJ (2009) The normalization model of attention.
\newblock Neuron 61: 168-185.
\bibAnnoteFile{reynoldsheeger2009}

\bibitem{hermannheeger2010}
Herrmann K, Montaser-Kouhsari L, Carrasco M, Heeger DJ (2010) When size
  matters: attention affects performance by contrast or response gain.
\newblock Nature Neuroscience 13: 1554-1559.
\bibAnnoteFile{hermannheeger2010}

\bibitem{leemaunsell2009}
Lee J, Maunsell JH (2009) A normalization model of attentional modulation of
  single unit responses.
\newblock PloS one 4: e4651.
\bibAnnoteFile{leemaunsell2009}

\bibitem{churchlandditterich}
Churchland AK, Ditterich J (2012) New advances in understanding decisions among
  multiple alternatives.
\newblock Current Opinion in Neurobiology 22: 920-926.
\bibAnnoteFile{churchlandditterich}

\bibitem{niwaditterich}
Niwa M, Ditterich J (2008) Perceptual decisions between multiple directions of
  visual motion.
\newblock Journal of Neuroscience 28: 4435-4445.
\bibAnnoteFile{niwaditterich}

\bibitem{bruntonbrody}
Brunton BW, Botvinick MM, Brody CD (2013) Rats and humans can optimally
  accumulate evidence for decision-making.
\newblock Science 340: 95-98.
\bibAnnoteFile{bruntonbrody}

\bibitem{tanner}
Tanner J W~P (1956) Theory of recognition.
\newblock Journal of the Acoustical Society of America 28: 882-888.
\bibAnnoteFile{tanner}

\bibitem{ashbybk}
Ashby FG (1992).
\newblock Multidimensional models of perception and cognition.
\bibAnnoteFile{ashbybk}

\bibitem{ashbytownsend}
Ashby FG, Townsend JT (1986) Varieties of perceptual independence.
\newblock Psychological review 93: 154-179.
\bibAnnoteFile{ashbytownsend}

\bibitem{swetsbirdsall}
Swets J, Birdsall TG (1956) The human use of information--iii: Decision-making
  in signal detection and recognition situations involving multiple
  alternatives.
\newblock IRE Transactions on Information Theory 2: 138-165.
\bibAnnoteFile{swetsbirdsall}

\bibitem{thomasolzak}
Thomas JP, Olzak LA (1992) Simultaneous detection and identification,
  Hillsdale, New Jersey: Lawrence Erlbaum Associates, Inc.
\newblock Multidimensional models of perception and cognition (Ed. F.G.Ashby).
  pp. 253-278.
\bibAnnoteFile{thomasolzak}

\bibitem{klein}
Klein SA (1985) Double-judgment psychophysics: problems and solutions.
\newblock Journal of the Optical Society of AmericaA, Optics and image science
  2: 1560-1585.
\bibAnnoteFile{klein}

\bibitem{olzakthomas}
Olzak LA, Thomas JP (1981) Gratings: why frequency discrimination is sometimes
  better than detection.
\newblock Journal of the Optical Society of AmericaA, Optics and image science
  71: 64-70.
\bibAnnoteFile{olzakthomas}

\bibitem{carandinichurchland}
Carandini M, Churchland AK (2013) Probing perceptual decisions in rodents.
\newblock Nature Neuroscience 16: 824-831.
\bibAnnoteFile{carandinichurchland}

\end{thebibliography}


\clearpage
\section*{Supporting Information: Figures and Tables}
\addcontentsline{toc}{section}{Supporting Information: Figures and Tables} 

\setcounter{figure}{0}
\makeatletter
\renewcommand{\thefigure}{S\@arabic\c@figure}

\begin{figure}[!h]
\begin{center}
\includegraphics[width=6.35in]{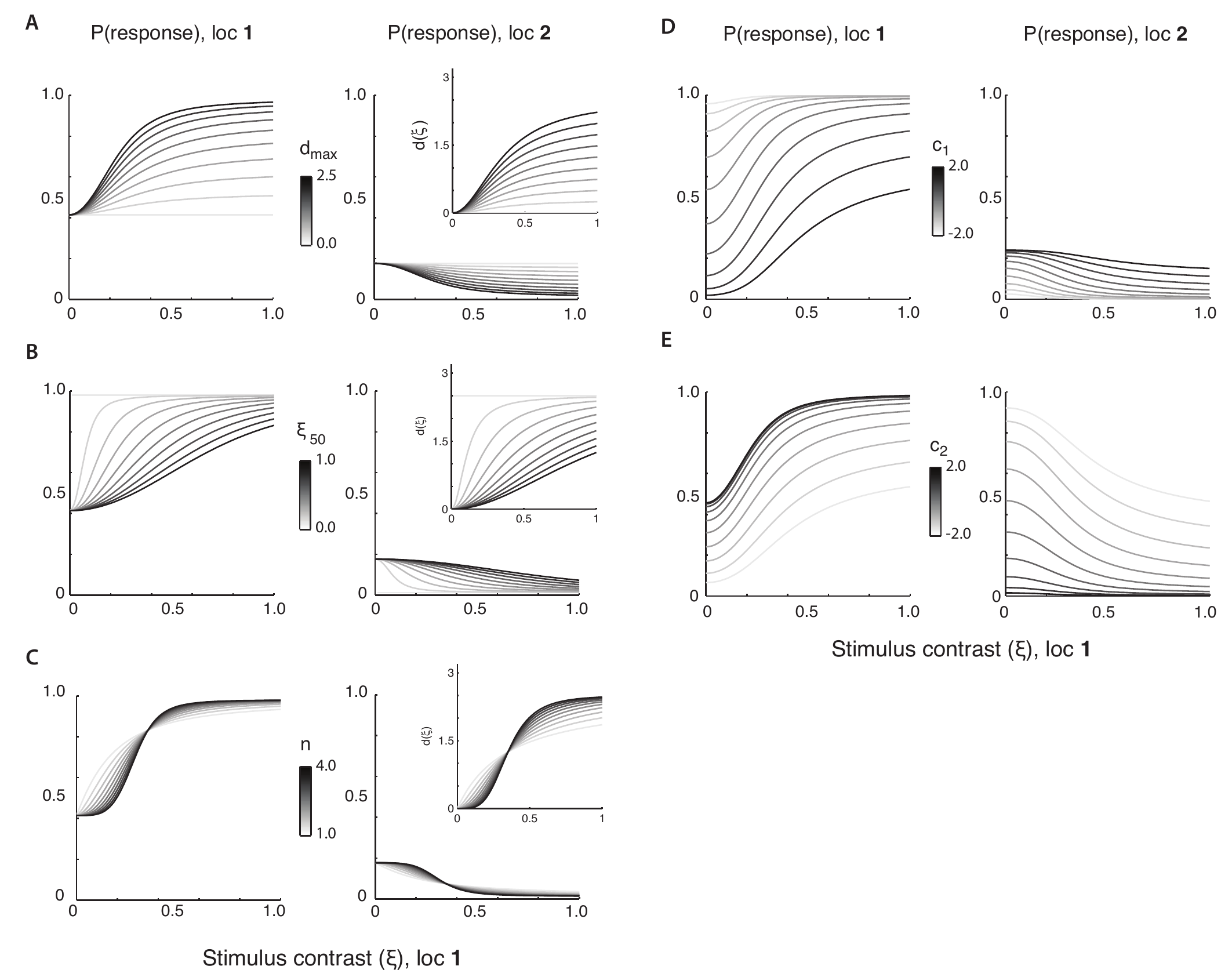}
\end{center}
\caption{
{\bf Effect of varying psychometric parameters and criteria on 2-ADC response probabilities}  
({\bf A}) ({\it Left}) Response probabilities $p(\xi)$ at location 1 as a function of stimulus contrast $\xi$ at location 1. The family of curves (light gray to black) correspond to increasing values of asymptotic sensitivity $d_{max}$ at location 1. ({\it Right}) Same as in left panel, but response probabilities at location 2 as a function of stimulus contrast at location 1. ({\it Inset}) Perceptual sensitivity $d$ as a function of stimulus contrast $\xi$ for increasing $d_{max}$ (scale parameter).
({\bf B}) Same as in (A), but response probabilities for increasing values of half-max contrast $\xi_{50}$ (shift parameter).
({\bf C}) Same as in (A), but response probabilities for increasing values of the exponent $n$ (slope parameter).
({\bf D}) Same as in (A), but response probabilities for increasing values of the criterion at location 1, $c_1$.
({\bf E}) Same as in (A), but response probabilities for increasing values of the criterion at location 2, $c_2$.
}
\label{fig_psyc}
\end{figure}

\makeatother

\clearpage
\includepdf[pages={1-3}, addtotoc={20,section,1,SI Tables,psyd}]{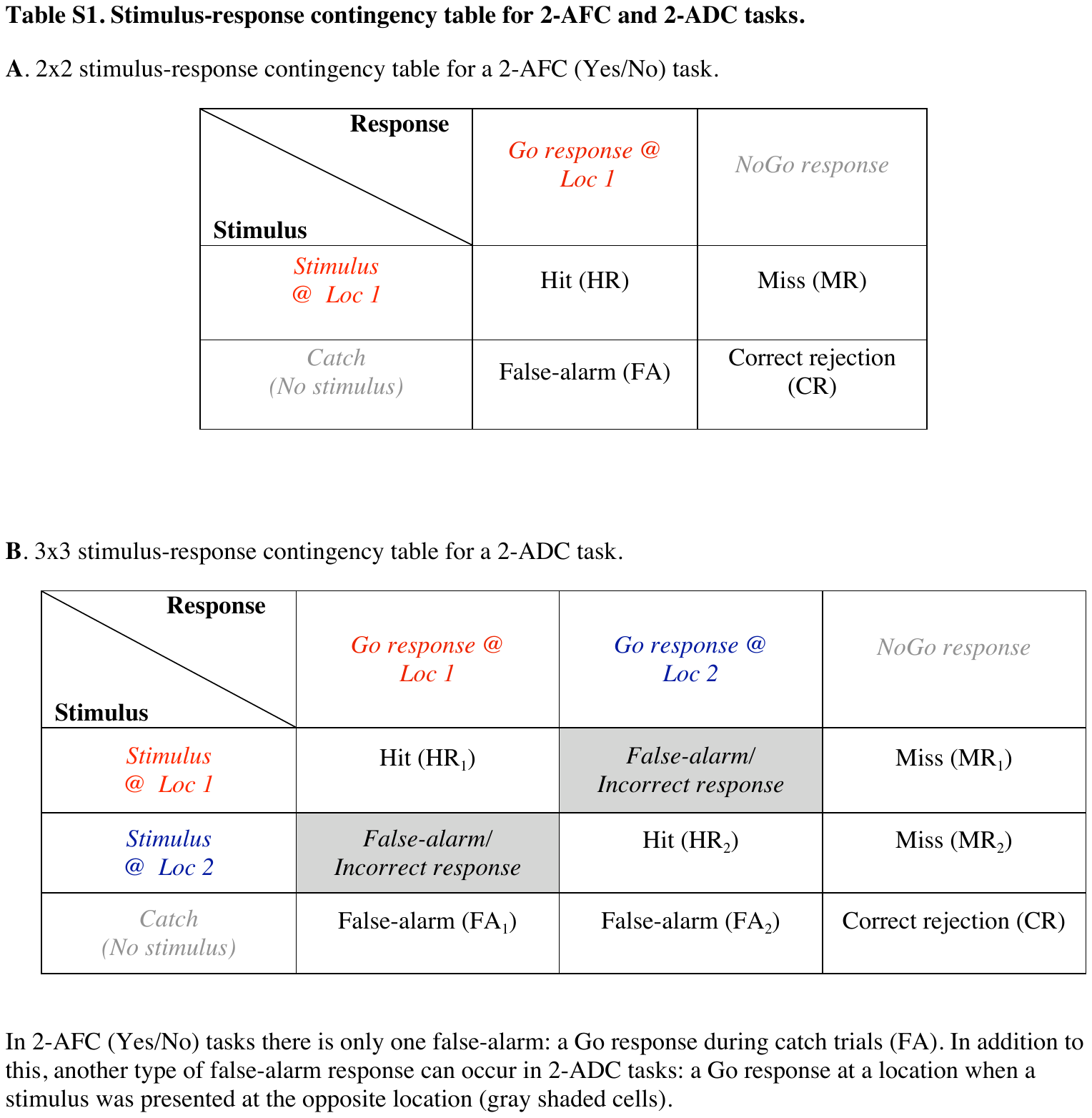}


\clearpage
\appendix
\section*{Supporting Information: Appendix}
\addcontentsline{toc}{section}{Supporting Information: Appendix}

\section{Proof of lemmas on the dependence of response probabilities on sensitivity and criteria}
\subsection{Proof of Lemma 1: Monotonic variation of the probability of a NoGo response with choice criteria in 2-ADC catch trials}

\noindent
\textbf{Assertion:} \textit{$p(Y=0 | X_1=0, X_2=0; c_1, c_2)$ or $p^0_0$ is a monotonically increasing function of both $c_1$ and $c_2$. Specifically, $p^0_0 = F_1(c_1) \ F_2(c_2)$.}

\vspace{11 pt}

\noindent
\textbf{Proof:}
The most straightforward proof for this lemma relies on the fact that NoGo responses in the 2-ADC task occur when each decision variable independently falls below the choice criterion at the respective location, and therefore, may be considered the joint outcome of two independent binary (Yes/No) decisions. Thus, the probability of a NoGo response when no target is presented at either location (correct rejection) $p^0_0$ factors into the product of the probability of NoGo response when no target is present (correct rejection) in each Yes/No decision. Thus,
\begin{eqnarray} 
p^0_0 & = & p(Y = 0 | X_1 = 0, X_2 = 0) \vert_{2-ADC} \nonumber \\ 
 & = & p(Y = 0 | X_1 = 0)\vert_{Yes/No} \  p(Y = 0 | X_2 = 0)\vert_{Yes/No} \nonumber \\
 & = & F_1(c_1) \ F_2 (c_2)
\end{eqnarray} 

For the analytical proof we proceed, as before. The condition for a NoGo response is that $\Psi$ falls below the criterion at both locations ($Y = 0, \mathrm{iff} \Psi_1 \leq c_1 \ \cap \ \Psi_2 \leq c_2 $). Thus, 
\begin{equation}
p(Y = 0 | X_1=0, X_2=0) =  p(\Psi_1 \leq c_1 \cap \ \Psi_2 \leq c_2 ) \\
\end{equation}

Upon substitution of the structural model, and noting that the noise distributions are assumed independent, this gives: 
\begin{eqnarray}
p(Y = 0 | X_1=0, X_2=0) & = & p(\varepsilon_1 \leq c_1 \cap \varepsilon_2 \leq c_2) \nonumber \\ 
 & = & p(\varepsilon_1 \leq c_1) p(\varepsilon_2 \leq c_2)  \nonumber \\
 & = & F_1(c_1) \ F_2(c_2)
\end{eqnarray}

Thus, the probability of a correct rejection in the 2-ADC task factors into the product of the 1-ADC correct-rejection probabilities. 
\begin{equation}
p^0_0 = F_1(c_1) \ F_2(c_2) \\
\end{equation}

As the $F_i$-s are positive, and monotonically increasing functions of their arguments, $p^0_0$ is a monotonically increasing function of $c_1$ and $c_2$.

\subsubsection*{Verification that $\sum_{i} p^i_0 = 1$}
The task specifies that the probabilities of response to the three locations are mutually exclusive and exhaustive. Below we perform a consistency check to ensure that these probabilities indeed sum to unity. Without loss of generality, we demonstrate this for the no stimulus condition $X_1=0, X_2=0$. The cases with a stimulus at either location may be identically derived by replacing the appropriate criterion by its difference with the respective sensitivity parameter, in other words, replace $c_i$ with $c_i-d_i$. Some of the results derived below will be used for later demonstrations.

We rewrite the equations of (\ref{eq:pi0a}) by changing the variable to integration to $e_i'$ where $e_i' = e_i - c_i, i \in \{1,2\}$. This gives:
\begin{eqnarray} 
p^1_0  & = &  \int_{0}^{\infty}  F_2(e_1' + c_2) \ f_1(e_1' + c_1) \ de_1' \label{eq:pi0b1} \\
p^2_0  & = &  \int_{0}^{\infty}  F_1(e_2' + c_1) \ f_2(e_2' + c_2) \ de_2' \label{eq:pi0b2} 
\end{eqnarray} 

Note that the change of variables altered the limits of integration such that these are no longer dependent on $c_i$. Next we integrate equation (\ref{eq:pi0b1}) by parts. 
\begin{eqnarray}
p^1_0  & = &  \left[F_2(e_1' + c_2) \ F_1(e_1' + c_1)\right]_{0}^{\infty} - \int_{0}^{\infty}  F_1(e_1' + c_1) \ f_2(e_1' + c_2) \ de_1' \\
 & = &  1 - (F_2(c_2) \ F_1(c_1)) - \int_{0}^{\infty}  F_1(e_1' + c_1) \ f_2(e_1' + c_2) \ de_1'  \label{eq:cvar}
\end{eqnarray}

where we have used the following identities of probability densities and cumulative densities: 
\begin{align}
d(F_i(x + k))& = f_i(x + k);& \int f_i(x + k) dx& = F_i(x + k);& \lim_{x \to \infty}F_i(x)& = 1 \label{eq:idpcdf}
\end{align}

The third term on the right hand side of equation (\ref{eq:cvar}), $ \int_{0}^{\infty}  F_1(e_1' + c_1) f_2(e_1' + c_2) de_1' $ is identical with $p^2_0$ of equation (\ref{eq:pi0b2}) (the $e_i'$ are interchangeable, as they are simply variables of integration). Thus, this equation may be written as:
\begin{eqnarray}
p^1_0 & = &  1 - (F_2(c_2) \ F_1(c_1)) - p^2_0 \\
1 - (p^1_0 + p^2_0) & = &  (F_2(c_2) \ F_1(c_1))
\end{eqnarray}

The left hand side of the above equation is nothing but $p^0_0$. Thus, we have verified that $\sum_{i} p^i_0 = 1$. 

\subsection{Proof of Lemma 2: Monotonic variation of the probabilities of Go responses with choice criteria in 2-ADC catch trials}

\noindent
\textbf{Assertion:} \textit{$p(Y=i| X_1=0, X_2=0; c_1, c_2)$ or $p^i_0$ is a monotonically decreasing function of $c_i$ and a monotonically increasing function of $c_j$.}

\vspace{11 pt}

\noindent
\textbf{Proof:} 

We reproduce equations (\ref{eq:pi0a}) here:
\begin{eqnarray}
p^1_0  & = &  \int_{c_1}^{\infty}  F_2(e_1 + c_2 - c_1) \ f_1(e_1) \ de_1 \\
p^2_0  & = &  \int_{c_2}^{\infty}  F_1(e_2 + c_1 - c_2) \ f_2(e_2) \ de_2 
\end{eqnarray}

With increasing $c_1$, $p^1_0$ has to decrease because: 
\begin{enumerate}[(i)]
\item The integrand ($F_2(e_1 + c_2 - c_1)$, specifically) decreases because $F_2$ is a monotonic function of its arguments ($e_1 + c_2 - c_1$ decreases)
\item The domain of integration ($c_1 \to \infty$) decreases as $c_1$ increases (the integrand is never negative)
\end{enumerate}
With increasing $c_2$, $p^1_0$ has to increase because the integrand increases ($F_2(e_1 + c_2 - c_1)$, specifically), and the domain of integration is unaffected by $c_2$. The (converse) effects of $c_1$ and $c_2$ on $p^2_0$ can be similarly argued.  

We substantiate this argument by quantifying the change in the response probabilities with choice criteria. This analysis is provided for the 2-ADC case only.

We begin with equations (\ref{eq:pi0b1}) and (\ref{eq:pi0b2}). 
\begin{eqnarray} 
p^1_0  & = &  \int_{0}^{\infty}  F_2(e + c_2) \ f_1(e + c_1) \ de \nonumber \\
p^2_0  & = &  \int_{0}^{\infty}  F_1(e + c_1) \ f_2(e + c_2) \ de \nonumber
\end{eqnarray} 

In the above, we have removed subscripts and the $'$ on the $e_i'$-s, as these are simply dummy variables of integration. Due to the symmetry of form, the proof of monotonicity for any one of $p^1_0$ or $p^2_0$ suffices. 

The assertion is proved in two parts: (i) we quantify the partial derivative of $p^1_0$ with respect to $c_1$, and show that it is negative  (${\partial p^1_0}/{\partial c_1} \le 0$), and (ii) we quantify the partial derivative of $p^1_0$ with respect to $c_2$, and show that it is positive  (${\partial p^1_0}/{\partial c_2} \ge 0$),

\noindent
Part (i): Differentiating $p^1_0$ with respect to $c_1$, we get:
\begin{equation}
\frac{\partial p^1_0}{\partial c_1} = \frac{\partial}{\partial c_1} \left[\int_{0}^{\infty}  F_2(e + c_2) \ f_1(e + c_1) \ de \right] 
\end{equation}

Applying the Leibniz integral rule:
\begin{eqnarray}
\frac{\partial p^1_0}{\partial c_1} & = & \int_{0}^{\infty} \frac{\partial ( F_2(e + c_2) \ f_1(e + c_1) )}{\partial c_1}  \ de \nonumber \\
 & = & \int_{0}^{\infty} F_2(e + c_2) \ \frac{\partial f_1(e + c_1)}{\partial c_1}  \ de 
\end{eqnarray}

Integrating by parts the expression on the right hand side:
\begin{equation}
\begin{split}
\frac{\partial p^1_0}{\partial c_1} & =  \left[ F_2(e + c_2) \ \frac{\partial F_1(e + c_1)}{\partial c_1} \right]_{0}^{\infty}  -  \int_{0}^{\infty} f_2(e + c_2) \ \frac{\partial F_1(e + c_1)}{\partial c_1}  \ de \\
 & = - F_2(c_2) \ \frac{\partial F_1(c_1)}{\partial c_1} -  \int_{0}^{\infty} f_2(e + c_2) \ \frac{\partial F_1(e + c_1)}{\partial c_1}  \ de \\ 
 & = - \left(F_2(c_2) \ f_1(c_1) +  \int_{0}^{\infty} f_2(e + c_2) \ f_1(e + c_1) \ de \right)
\end{split}
\end{equation}

\noindent
where we have used the identities from (\ref{eq:idpcdf}).

We note that both terms inside the parentheses on the right hand side are positive. The first is a product of a CDF and a PDF, the second is an integral over a function that is a product of two PDF-s, and hence never negative. Now taking the negative sign into consideration, it is clear that ${\partial p^1_0}/{\partial c_1}$ is negative for all $c_1$. 

\noindent
Part (ii): Similarly, differentiating $p^1_0$ with respect to $c_2$, we get:
\begin{equation}
\frac{\partial p^1_0}{\partial c_2} = \frac{\partial}{\partial c_2} \left[\int_{0}^{\infty}  F_2(e + c_2) \ f_1(e + c_1) \ de \right] 
\end{equation}

Again applying the Leibniz integral rule:
\begin{eqnarray}
\frac{\partial p^1_0}{\partial c_2} & = & \int_{0}^{\infty} \frac{\partial ( F_2(e + c_2) \ f_1(e + c_1) )}{\partial c_2}  \ de \nonumber \\
 & = & \int_{0}^{\infty} \frac{\partial F_2(e + c_2)}{\partial c_2} \ f_1(e + c_1) \ de \nonumber \\
 & = & \int_{0}^{\infty} f_2(e + c_2) \ f_1(e + c_1) \ de 
\end{eqnarray}

The integrand on the right hand side is a product of two probability density functions, and is positive for all values of $e$ (strictly positive if $f_1$ and $f_2$ are supported over $[0, \infty]$). Thus, ${\partial p^1_0}/{\partial c_2}$ is positive for all $c_2$. 

\subsection{Proof of Lemma 3: Monotonic variation of response probabilities with perceptual sensitivities in 2-ADC stimulus trials}

\textbf{Assertion:} \textit{$p(Y=i| X_i, X_j; d_i, d_j)$ or $p^i_j$ is a monotonic function of $d_i$ and $d_j (i,j \in \{1, 2\})$.}

\vspace{11 pt}

\noindent
\textbf{Proof:} 
We reproduce part of equation system (\ref{eq:sixindep3}) here for reference.
\begin{align}  
p(Y = i | X_i, X_j)& = \int_{c_i - d_i X_i}^{\infty}  F_j(e + d_i X_i - d_j X_j - c_i + c_j) \ f_i(e) \ de \\ \nonumber
& \footnotesize{i,j \in \{1,2\}, i \neq j} 
\end{align}

\noindent
where we have dropped the subscript from $e_i$ (a variable of integration). 

With increasing $d_i$, $p(Y = i | X_i=1)$ or $p^i_i$ has to increase because: 
\begin{enumerate}[(i)]
\item The integrand ($F_j(e + d_i - c_i + c_j)$, specifically) increases because $F_j$ is a monotonic function of its arguments ($d_i$ increases)
\item The domain of integration ($c_i - d_i \to \infty$) increases as $d_i$ decreases (the integrand is never negative)
\end{enumerate}

With increasing $d_j$, $p(Y = i | X_j=1)$ or $p^i_j$ has to decrease because: 
\begin{enumerate}[(i)]
\item The integrand ($F_j(e - d_j - c_i + c_j)$, specifically) increases because $F_j$ is a monotonic function of its arguments (-$d_j$ decreases)
\item The domain of integration is unaffected by $d_j$.
\end{enumerate}

We substantiate this argument by quantifying the change in the response probabilities with perceptual sensitivity. This analysis is provided for the 2-ADC case only.

We rewrite the above equations with the following transformation $e' = e - c_i + d_i X_i$; as before, the idea is to eliminate the occurrence of criterion and sensitivity in the limits of integration. The system may then be rewritten as: 
\begin{align}  
p(Y = i | X_i, X_j)& = \int_{0}^{\infty}  F_j(e' + c_j - d_j X_j) \ f_i(e' + c_i - d_i X_i) \ de' \\ \nonumber
& \footnotesize{i,j \in \{1,2\}, i \neq j} \label{eq:sixindep3A}
\end{align}

\noindent
In the following, we replace the notation for variable $e'$ with $e$. Computing the partial derivative of system (\ref{eq:sixindep3}) with respect to $d_i$:
\begin{equation}
\begin{split}
\frac{\partial p(Y = i | X_i, X_j)}{\partial d_i}& = \frac{\partial}{\partial d_i} \left(\int_{0}^{\infty}  F_j(e + c_j - d_j X_j) \ f_i(e + c_i - d_i X_i) \ de \right) \\
&= \int_{0}^{\infty} \frac{\partial}{\partial d_i} \ \left(F_j(e + c_j - d_j X_j) \ f_i(e + c_i - d_i X_i)\right) \ de \\
&= \int_{0}^{\infty}  \ F_j(e + c_j - d_j X_j)  \left(\frac{\partial f_i(e + c_i - d_i X_i)} {\partial d_i}\right) \ de
\end{split}  
\end{equation}

Integrating by parts, and noting that $\frac{\partial}{\partial d_i}(d_k X_k) = \delta_{ik} X_k$, we have: 
\begin{equation}
\begin{split}
\frac{\partial p(Y = i | X_i, X_j)}{\partial d_i}& = \left[ F_j(e + c_j - d_j X_j) \frac{\partial F_i(e + c_i - d_i X_i)}{\partial d_i} \right]_{0}^{\infty}  \\ 
 & \ \ \ \ \ - \int_{0}^{\infty}  f_j(e + c_j - d_j X_j) \ \frac{\partial F_i(e + c_i - d_i X_i)}{\partial d_i} \ de\\
&= - F_j(c_j - d_j X_j) \ f_i(c_i - d_i X_i) (-X_i) \\ 
 & \ \ \ \ \ - \int_{0}^{\infty}  f_j(e + c_j - d_j X_j) \  f_i(e + c_i - d_i X_i) (-X_i) \ de  \\
&= X_i \ \{ F_j(c_j - d_j X_j) \ f_i(c_i - d_i X_i) \\ 
& \ \ \ \ \  + \int_{0}^{\infty}  f_j(e + c_j - d_j X_j) \  f_i(e + c_i - d_i X_i) \ de\}
\end{split}  
\end{equation}

The right hand side is clearly positive being the sum of two positive terms: the product of a CDF and PDF, and the integral of the product of two PDF-s, respectively ($X_i \geq 0, X_i \in \{0, 1\}$). Thus, 
\begin{equation}
\frac{\partial}{\partial d_i}  p(Y = i | X_i, X_j) \geq 0 \ \ \ \ \ \ \ \forall \ d_i \label{eq:ap2a}
\end{equation} 

Specifically, $\frac{\partial}{\partial d_i} p(Y = i | X_i=0, X_j=1) = \frac{\partial p^i_j}{\partial d_i} = 0$, which confirms the intuition that the response probability at location $i$ to a stimulus at the opposite location $j$ does not depend on the sensitivity at location $i$. 

Similarly, computing the partial derivative of system (\ref{eq:sixindep3}) with respect to $d_j$:
\begin{equation} \label{eq:ap1}
\begin{split}
\frac{\partial p(Y = i | X_i, X_j)}{\partial d_j} &= \int_{0}^{\infty} \left( \frac{\partial F_j(e + c_j - d_j X_j)} {\partial d_j} \right) \ f_i(e + c_i - d_i X_i) \ de  \\
 &=  \int_{0}^{\infty} f_j(e + c_j - d_j X_j) (-X_j) f_i(e + c_i - d_i X_i) \ de  \\
 &= -X_j  \int_{0}^{\infty} f_j(e + c_j - d_j X_j)  f_i(e + c_i - d_i X_i) \ de 
\end{split}  
\end{equation}

The right hand side of (\ref{eq:ap1}) is clearly negative, as the integrand (product of two PDFs) is positive and $X_j \in \{0, 1\}$. Thus,
\begin{equation} 
\frac{\partial}{\partial d_j}  p(Y = i | X_i, X_j) \leq 0 \ \ \ \ \ \ \ \forall \ d_j \label{eq:ap2b}
\end{equation} 

(\ref{eq:ap2a}) and (\ref{eq:ap2b}) complete the proof.

\subsection{Proof of Lemma 4: One-to-one correspondence of m-ADC choice criteria}
\textbf{Assertion:} \textit{Given a set of response probabilities $p^r_0=\mathcal{P}^r_0, r \in \{0, \ldots, m+1\}$, and any solution set $C = \{c_j: j \in \{0, \ldots, m+1\}\}$ comprising ordered sets of choice criteria satisfying the system (\ref{eq:madc_cth}). There is a one-to-one mapping between any choice criterion $c_i$ and its complement set $C'_i = \{c_j: j \in \{0, \ldots, m+1\}, j \neq i\}$}. 

\vspace{11 pt}

\noindent
\textbf{Proof:} 
The proof proceeds in two steps, first demonstrating the mapping $\phi: C'_i \mapsto c_i$, and then its inverse $\phi: c_i \mapsto C'_i$. 

First, consider the probability $p^0_0=\mathcal{P}^0_0$. A given choice criterion $c_i, i \in \{0, \ldots, m+1\}$ can be expressed in terms of the remaining criteria in the following way.
\begin{align}
\mathcal{P}^0_0 &= \prod_{j=1}^{m+1} F_j(c_j) 	\\
c_i &= F_i^{-1}(\frac{\mathcal{P}^0_0}{ \prod_{j=1, j \neq i}^{m+1} F_j(c_j) }) 
\end{align}

where $F_i$ is invertible because it is monotonic (being a cumulative density function). Given a particular $p^0_0=\mathcal{P}^0_0$, and a solution set of $m$ criteria $\{c_j: j \in {0, \ldots, m+1}, j \neq i\}$ the remaining criterion $c_i$ is uniquely determined, thus demonstrating the mapping $\phi: C'_i \mapsto c_i$.

Next, consider the set of probabilities $p(Y = i | \ \lVert \boldsymbol{X} \rVert_{1}=0) = \mathcal{P}^i_0$. From system (\ref{eq:madc_cth}), these can be written as:
\begin{align}
\mathcal{P}^i_0 &=  \int_{c_i}^{\infty} \prod_{k=1, k \neq i}^{m+1}  F_k(e_i - c_i + c_k) \ f_i(e_i) \ de_i 
\end{align}

With the substitution $e=e_i-c_i$, and following some algebra (see \ref{eq:pi0b1}), this set of equations can be rewritten as: 
\begin{align}
\mathcal{P}^i_0 &=  \int_{0}^{\infty} \prod_{k=1, k \neq i}^{m+1}  F_k(e + c_k) \ f_i(e + c_i) \ de \label{eq:masubs}
\end{align}

Let C be a set of criteria $\{c_j: j \in \{0, \ldots, m+1\}\}$ that satisfies this equation. Let us assume that one of the criteria in this set, say $c_{m+1}$ (without loss of generality) has a known value.  

Define the following functions (for $i \in {0, \ldots, m}$). 
\begin{align}
F_{\mu}(e; c_{m+1}) &= \left[ F_{m+1}(e + c_{m+1}) \right]^\frac{1}{m} \\
G_i(e+c_i; c_{m+1}) &= F_i(e+c_i) \ F_{\mu}(e; c_{m+1}) 
\end{align}

We note that both $F_{\mu}$ and $G_i$ are parameterized by $c_{m+1}$. $F_{\mu}$, the $m$-th root of a cumulative density function, and $G$, the product of $F_{\mu}$ and $F_i$ are both monotonic, continuous functions, and it is easy to see that $\lim \limits_{e \to -\infty} G_i = 0; \lim \limits_{e \to +\infty} G_i = 1$. Thus, $G_i$ is itself a cumulative density function with the following probability density:
\begin{align}
g_i(e+c_i; c_{m+1}) &= \frac{\partial G}{\partial e} \\ 
&= F_i(e+c_i) \ \frac{\partial F_{\mu}(e; c_{m+1})}{\partial e} +  F_{\mu}(e; c_{m+1}) \ f_i(e+c_i) 
\end{align}
 
Now, let us consider the following system of equations:
\begin{align}
\mathcal{Q}^i &= \int_{0}^{\infty} \prod_{k=1, k \neq i}^{m} G_k(e+c_k; c_{m+1}) \ g_i(e+c_i; c_{m+1}) \ de 
\end{align}

With some algebra, we can show that $\mathcal{Q}^i = \mathcal{P}^i_0 + (\mathcal{P}^{m+1}_0 / m)$.

By the induction hypothesis for m-equations, given a set of $q^i$-s, and the parameter $c_{m+1}$, all of the $c_k$-s are uniquely determined. Because $c_{m+1}$ was an arbitrarily chosen criterion, the result can be generalized as follows: given a set of $\mathcal{Q}^i$-s, and any choice criterion $c_i$, all of the other choice criteria in $C'_i = \{c_j: j \in \{0, \ldots, m+1\}, j \neq i\}$-s are uniquely determined, thus demonstrating the inverse mapping $\phi^{-1}: c_i \mapsto C'_i$.
 
Thus, we prove the one-to-one mapping among any one choice criterion, and the remaining criteria, $c_i \leftrightarrow C'_i$.

\subsection{Proof of Lemma 5: Direct variation among all criteria in the m-ADC task}
\textbf{Assertion:} \textit{Given a set of response probabilities $\mathcal{P}^i_0, i \in {1, \ldots, m+1}$ and the set of all solution sets $\{C^k = \{{c_j}^k: j \in {0, \ldots, m+1}\}\}$ comprising ordered sets of choice criteria satisfying the system (\ref{eq:madc_cth}). For any two solution sets $C^1$ and $C^2$ every pair of corresponding elements $({c_j}^1, {c_j}^2)$ obeys the same order relation i.e. if any ${c_i}^1 \gtrless {c_i}^2$ then every ${c_j}^1 \gtrless {c_j}^2, i,j \in {0, \ldots, m+1}$.}

\vspace{11 pt}

\noindent
\textbf{Proof:}  
Let $C^1= \{{c_j}^1: j \in {0, \ldots, m+1}\}$ be a solution set of criteria satisfying the system (\ref{eq:madc_cth}), and let $C^2= \{{c_j}^2: j \in {0, \ldots, m+1}\}$  be another, distinct (not identical) solution set. Also let all choice criteria from set $C^1$, except that corresponding to choice i ($c_i$), be greater (or lesser) in value than the corresponding criteria in set $C^2$. We demonstrate that in this case, the criterion $c_i$ in set $C^1$ must also be greater (or lesser) in value than the corresponding criterion in set $C^2$.

Given the probability of a Go response to choice $i$ during catch trials, this can be written as (\ref{eq:masubs}):
\begin{align}
\mathcal{P}^i_0 &= \int_{0}^{\infty} \prod_{j=1, j \neq i}^{m+1}  F_j(e + {c_j}^1) \ f_i(e + {c_i}^1) \ de \\	\nonumber
&=  \int_{0}^{\infty} \prod_{j=1, j \neq i}^{m+1}  F_j(e + {c_j}^2) \ f_i(e + {c_i}^2) \ de 			\label{eq:pi0masubs}
\end{align}

Note that if ${c_j}^1 \gtrless {c_j}^2$,
\begin{align}
\prod_{j=1, j \neq i}^{m+1}  F_j(e + {c_j}^1) &\gtrless \prod_{j=1, j \neq i}^{m+1}  F_j(e + {c_j}^2) \ \ \  \forall e
\end{align}
 
as the $F_j$-s are monotonically increasing functions of their arguments. Hence, for the right hand sides of equation (\ref{eq:pi0masubs}) to be equal to each other (and each equal to $\mathcal{P}^i_0$) ${c_i}^1 \gtrless {c_i}^2$. The latter result is confirmed by inspecting the integrands of equation (\label{eq:pi0masubs}), and is also evident from the following lemma. 

\begin{quote}
\textbf{Lemma 8} \ \ The response probability $p^i_0$ is a strictly (monotonically) increasing function of $c_j$ and a strictly (monotonic) decreasing function of $c_i$.
\end{quote}

The lemma is proved in a subsequent section. Thus, if every ${c_j}^1 \gtrless {c_j}^2, j \in {0, \ldots m+1}, j\neq i$ then, ${c_i}^1 \gtrless {c_i}^2$.     

However, we have just shown that there is a one-to-one mapping between each $c_i$ and its complement set $C'_i = \{c_j: j \in \{0, \ldots, m+1\}, j \neq i\}$. Thus, the converse statement must also hold: that is, if ${c_i}^1 \gtrless {c_i}^2$, then every ${c_j}^1 \gtrless {c_j}^2, j \in {0, \ldots m}, j \in \{0, \ldots, m+1\}, j\neq i$. This completes the proof.   

\subsection{Proof of Lemma 6: Inverse variation among at least a pair of criteria in the m-ADC task}
\textbf{Assertion:} \textit{Given a set of response probabilities $ \mathcal{P}^0_0 $ and the set of all solution sets $\{C^k = \{{c_j}^k: j \in {0, \ldots, m+1}\}\}$ comprising ordered sets of choice criteria satisfying the system (\ref{eq:madc_cth}). For any two solution sets $C^1$ and $C^2$ at least one pair of corresponding elements $({c_j}^1, {c_j}^2)$ differs in its order relation i.e. if any ${c_i}^1 \gtrless {c_i}^2,  i \in {0, \ldots, m+1}$ then at least one ${c_j}^1 \lessgtr {c_j}^2$.}

\vspace{11 pt}

\noindent
\textbf{Proof:} 
Let $C^1= \{{c_j}^1: j \in {0, \ldots, m+1}\}$ and $C^2= \{{c_j}^2: j \in {0, \ldots, m+1}\}$ be distinct solution sets of criteria satisfying the system (\ref{eq:madc_cth}). Also let any one choice criterion from set $C^1$, corresponding to choice i ($c_i$), be greater in value than the corresponding criterion in set $C^2$ i.e. ${c_i}^1 > {c_i}^2$. 

We prove the result by contradiction. Assume that \textit{none} of the other criteria in set $C^1$ is lesser than the corresponding criteria in set $C^2$. In other words, every ${c_j}^1 \geq {c_j}^2, j \in {0, \ldots m+1}$.  

Given the probability of a NoGo response during catch trials, this can be written as:
\begin{align}
\mathcal{P}^0_0 &= \prod_{j=1}^{m+1} F_j({c_j}^1) &=  \prod_{j=1}^{m+1} F_j({c_j}^2) \label{eq:p00masubs}
\end{align}

The functions $F_j$ are monotonic functions of their arguments. If every ${c_j}^1 \geq {c_j}^2$ equality of the right hand side expressions holds only if ${c_j}^1 = {c_j}^2$, which violates the assumption that $C^1$ and $C^2$ are non-identical sets. Thus, if any one ${c_i}^1 > {c_i}^2$, the assumption that \textit{none} of the other criteria in set $C^1$ is lesser than the corresponding criteria in set $C^2$ leads to a contradiction. Hence, if any ${c_i}^1 > {c_i}^2,  i \in {0, \ldots, m+1}$ then at least one criterion in set $C_1$ has to be lesser than the corresponding criterion in set $C_2$. 

It is easy to see that the converse is also true, i.e. if any ${c_i}^1 < {c_i}^2,  i \in {0, \ldots, m+1}$ then at least one criterion in set $C_1$ has to be greater than the corresponding criterion in set $C_2$. This completes the proof. 

\subsection{Proof of Lemma 7: Monotonic variation of m-ADC response probabilities with perceptual sensitivity}
\textit{Assertion:} The response probability $p^i_j$ is a strictly monotonic (increasing) function of $d_i$ and a strictly monotonic (decreasing) function of $d_j$.

\vspace{11 pt}

\noindent
\textbf{Proof:} 
We reproduce the system of equations (\ref{eq:maufc}) for reference:
\begin{align}
p^i_j &=  \int_{c_i - d_i X_i}^{\infty} \prod_{k=1, k \neq i}^{m}  F_k(e_i + d_i X_i - d_k X_k - c_i + c_k) \ f_i(e_i) \ de_i 
\end{align}

Consider the probability of response to location $i$ when the stimulus is presented at the same location ($X_i = 1, X_k = 0 \ \forall \ k \neq i$).
\begin{align}
p^i_i &=  \int_{c_i - d_i}^{\infty} \prod_{k=1, k \neq i}^{m}  F_k(e_i + d_i - c_i + c_k) \ f_i(e_i) \ de_i 
\end{align}

With increasing $d_i$, the response probability $p^i_i$ has to increase, as the integrand increases with $d_i$ (each $F_k$ is a monotonically increasing function of its argument), and the integration (positive integrand) occurs over a larger domain ($c_i - d_i$ decreases). 

Next, consider the probability of response to location $i$ when the stimulus is presented at location $j, j \neq i$ ($X_j = 1, X_k = 0 \forall k \neq j$).
\begin{align}
p^i_j &=  \int_{c_i}^{\infty} \prod_{k=1, k \neq i,j}^{m}  F_k(e_i - c_i + c_k) F_j(e_i - d_j - c_i + c_j) \ f_i(e_i) \ de_i 
\end{align}
Again, it is apparent that with increasing $d_j$ the response probability $p^i_j$ has to decrease, as the integrand ($F_j(e_i - d_j - c_i + c_j)$, specifically) decreases with increasing $d_j$ (the domain of integration is unaffected by $d_j$). 

This completes the proof. The assertion can be proved analytically by quantifying the change in response probabilities with perceptual sensitivity following a procedure similar to that of Lemma 3.

\subsection{Proof of Lemma 8: Monotonic variation of m-ADC response probabilities with choice criteria}
\textit{Assertion:} The response probability $p^i_0$ is a strictly monotonic (decreasing) function of $c_i$ and a strictly monotonic (increasing) function of $c_j$.

\vspace{11 pt}

\noindent
\textbf{Proof:} 
Consider the probability of response to location $i$ when the no stimulus is presented ($X_k = 0 \forall k$).
\begin{align}
p^i_0 &=  \int_{c_i}^{\infty} \prod_{k=1, k \neq i}^{m}  F_k(e_i - c_i + c_k) \ f_i(e_i) \ de_i 
\end{align}

With increasing $c_i$, the response probability $p^i_0$ has to decrease as the integrand decreases with $c_i$ (each $F_k$ a monotonically decreases with $c_i$), and the integration (positive integrand) occurs over a smaller domain. Similarly, with increasing $c_j$, $p^i_0$ has to increase as the integrand  ($F_j(e_i - c_i + c_j)$, specifically) increases with $c_j$.

This completes the proof. The analytical demonstration by quantifying the changes in response probabilities with choice criteria follows a procedure similar to that of Lemma 2.


\section{The m-AFC model with bias as a special case of the m-ADC model}

In the m-ADC model, if the decision variable never falls below the criterion at any location, the subject never provides a NoGo response. This can be achieved by setting the criteria to very low (large negative) values. In this case the m-ADC model is identical with an m-AFC model, assuming no catch trials are incorporated in the task design.

Formally, the m-ADC model reduces to the m-AFC model in the limit $c_i \to -\infty$ (Figure 7A vs. 7B) while preserving the difference between every pair of criteria $c_i - c_j = b_{ij}$, which we term the bias for location $j$ relative to location $i$. Applying this limit to equation (\ref{eq:maufc}):
\begin{eqnarray}
p(Y = i | \mathbf{X}) & = & \int_{-\infty}^{\infty} \prod_{k, k \neq i}  F_k(e_i + d_i X_i - d_k X_k - b_{ik}) \ f_i(e_i) \ de_i \label{eq:mafc}
\end{eqnarray}

These equations describe a recently developed m-AFC model formulation that incorporates bias \cite{decarlo}. Thus, the m-ADC model is a more general form of the m-AFC model and the analytical results we have proved above for the m-ADC model are all valid for the m-AFC model (with bias) as well.

\end{document}